\documentclass[a4paper]{jpconf}
\usepackage{graphicx,amssymb}
\usepackage{cite}

\def\ket{\rangle}
\def\bra{\langle}

\def\epsd{\epsilon_d}

\def\dA{d_1}
\def\dB{d_2}

\def\epsEP{\bar{\epsilon}_A}
\def\eEP{\bar{E}_A}
\def\lamEP{\bar{\lambda}_{A}}
\def\delEP{\Delta_{EP}}

\def\lam{\lambda}
\def\lamB{\bar{\lambda}_{B}}

\newcommand{\beq}{\begin{equation}}
\newcommand{\eeq}{\end{equation}}
\newcommand{\beqa}{\begin{eqnarray}}
\newcommand{\eeqa}{\end{eqnarray}}

\begin{document}
\title{Characteristic influence of exceptional points in quantum dynamics}

\author{Savannah Garmon$^{1,2}$, Takafumi Sawada$^1$, Kenichi Noba$^1$, Gonzalo Ordonez$^3$}

\address{$^1$ Department of Physical Science, Osaka Prefecture University, Gakuen-cho 1-1, Sakai 599-8531, Japan}
\address{$^2$ Institute of Industrial Science, University of Tokyo, Kashiwa 277-8574, Japan}
\address{$^3$ Department of Physics and Astronomy, Butler University, Gallahue Hall, 4600 Sunset Avenue, Indianapolis, Indiana 46208, USA}

\ead{sgarmon@p.s.osakafu-u.ac.jp}


\begin{abstract}
We review some recent work on the occurrence of coalescing eigenstates at exceptional points in non-Hermitian systems and their influence on physical quantities.  We particularly focus on quantum dynamics near exceptional points in open quantum systems, which are described by an outwardly Hermitian Hamiltonian that gives rise to a non-Hermitian effective description after one projects out the environmental component of the system.  We classify the exceptional points into two categories: those at which two or more resonance states coalesce and those at which at least one resonance and the partnering anti-resonance coalesce (possibly including virtual states as well), and we introduce several simple models to explore the dynamics for both of these types.  In the latter case of coalescing resonance and anti-resonance states, we show that the presence of the continuum threshold plays a strong role in shaping the dynamics, in addition to the exceptional point itself.  We also briefly discuss the special case in which the exceptional point appears directly at the threshold.
\end{abstract}


\section{Introduction}
\label{sec:intro}


The traditional assumption of Hermiticity in quantum theories has enabled remarkable success by providing a key element in a formalism that is mathematically elegant and which yields excellent agreement with experiment.  However, in recent decades, many researchers have found it useful to introduce non-Hermitian elements into physical theories for a variety of reasons \cite{HNPRL96,HNPRB97,BB98,MostaJMP02,BBJ02,RDM05,MECM08,Rotter_review,Moiseyev-Book,RotterBird,GGH15,Konotop_review,PTBook,BenderBook}.  Most commonly, this has been done to describe energy lost to or acquired from the surrounding environment, which has often been described in terms of coupled-mode theory.  Under this kind of formalism, one assumes the most important physical properties of a given system can be described in terms of a few principal modes, one or more of which may have complex energy.  The imaginary parts of the complex energies can then be associated with energy sinks or sources from the surrounding environment.  This approach has been very successful in describing the physics of {\it exceptional points} (EPs) \cite{Kato,Berry04,Heiss12}, at which two or more eigenstates coalesce while their respective eigenvalues become degenerate, particularly in optics \cite{Alu_EP_review}.  Numerous interesting features of the exceptional points have been revealed both theoretically and in experiments, such as intrinsic chirality \cite{HeissChiral,RichterChiral} and modifications of the usual adiabatic properties when they are encircled \cite{GMM13,Rotter15,HeissNat,WunnerEncircle,RotterNat,XuNat,Liu21}.  Enhanced parametric sensitivity in the vicinity of the EPs may also lead to interesting applications \cite{Wiersig14,Chen,Stone19,ChristoEP3,WiersigEP3}.  From this point of view, some researchers have questioned if environmental influences and dissipation might better be viewed as a resource than a limitation.

However, in a deeper sense, there has always existed a tension between the strict requirement of Hermiticity and the actual implementation of the theory since the earliest days of quantum mechanics.  This is evident from the fact that quantum mechanics demands that energy eigenvalues must be real, yet the exponential decay width in well-established processes such as atomic relaxation is obtained from the negative imaginary part of a complex eigenvalue, which is the so-called {\it resonance} eigenvalue.  This `quirk' of the theory has generally been dealt with quite differently by different researchers.  Since the resonance eigenvalue is usually obtained by an analytic continuation of the original Hermitian theory into the complex energy plane, some have waved the issue away as being little more than a mathematical trick.  But others have elected to embrace the role of the resonance in the theory, for example by viewing it as a generalized eigenstate that resides outside of the usual Hilbert space (that only admits $L^2$-normalizable solutions with real eigenvalues) \cite{Moiseyev-Book,Nakanishi,SCG78,BohmBook,PPT91,Gadella11,Madrid12,Hatano_eff,TGKP16,KGTP17}.  Or, a more recent work \cite{HO14} has proposed a formalism in which the resonance state and bound state are treated on precisely equal footing (albeit with the inclusion of two other generalized eigenstates: the partner anti-resonance, with positive imaginary part of the eigenvalue, as well as anti-bound or virtual states that have real eigenvalue but are non-normalizable in the usual sense) \cite{SHO11,GO17,OH17A,OH17B}.

A clear advantage to this approach is that the resonance appears spontaneously due to the interaction between the system and the surrounding environment, rather than simply being presupposed as in the coupled mode theory.  More precisely, under this approach, one describes the environment surrounding the primary system of interest in terms of an energy continuum.  The details of the energy continuum are determined by the microscopic degrees of freedom associated with a particular environment.  This includes the existence of a lower threshold (equivalent to a band edge or waveguide cutoff mode) below which the continuum ends and the usual bound state eigenvalues (if any) would appear (the continuum might or might not also have an upper threshold).  The continuum also has a built-in density of states, describing how many modes a given level within the continuum can accommodate.  Under a number of situations, either the threshold or the density of states (or both) might strongly influence the dynamical properties of a given system.  Indeed, since the resonance emerges in this picture precisely due to the interaction of the discrete system modes with the continuum, the resonance generally first appears in the vicinity of the threshold.  The point in the parameter space at which the resonance appears is precisely an exceptional point, which some authors have therefore interpreted as a kind of spectral phase transition at which the dynamics are abruptly modified \cite{PPT91,TGKP16,OH17A,OH17B,DBMP08,GRHS12}  Further, since the exceptional point at which the resonance arises lies close to the threshold, one would expect the threshold itself could strongly influence the dynamics in this case, which has been confirmed by two of the present authors in Ref. \cite{GO17}.  We review this work below.

In this paper, we refer to models that incorporate both discrete and continuous spectral elements as examples of {\it open quantum systems}. 
We note that in the context of cavity quantum electrodynamics, this microscopic description of the environment in terms of a continuum incorporating a well-defined density of states is often referred to as a {\it structured reservoir} \cite{NBL99,LNNB00,DG03,LonghiPRA06,GNHP09,LCP14,GTC2D1,GTC2D2,SZMG17,SXCY19}.  
We also comment that the coupled-mode theory described above can be viewed as a kind of macroscopic approximation over the microscopic degrees of freedom written into the structured reservoir, which explains why the coupled-mode theory has been particularly successful in the classical context.

In the following, we review some recent work on the dynamics near exceptional points from the microscopic perspective, particularly Ref. \cite{GO17} (see also \cite{GOH21}).  We note a disadvantage to the microscopic framework is that deriving the dynamics in the context of the structured reservoir requires significantly more work than in the coupled-mode theory.  However, an obvious advantage to this approach is that it should provide a more accurate description of the dynamics in certain truly quantum settings.  A further advantage we will find is that the dynamics are richer in the case that we preserve the microscopic degrees of freedom in the system.

It is worth mentioning that in this work, we only deal with the problem at the Hamiltonian level.  Some recent works have emphasized that a more complete description can be obtained by considering the Liouvillian formalism, which can describe quantum jumps and other effects \cite{BreuPet,HatanoLindblad,LEP,Haack}  We comment on this point further in the conclusion.

In the next section, we introduce a quick categorization scheme for two types of exceptional points in open quantum system, before outlining the remainder of the paper.


\section{Categorization scheme for two-level exceptional points}
\label{sec:EP.cat}

Here we briefly outline a convenient categorization scheme for the exceptional points as originally introduced in Refs \cite{GO17,GGH15}.

Let us begin by denoting an exceptional point at which $N$ eigenstates coalesce as an EPN.  Then in the simplest case of the EP2, there are two further natural subcategories.  First, an exceptional point at which two virtual states with real eigenvalues coalesce before forming a resonance/anti-resonance pair is defined as an EP2A.  As mentioned in the Introduction, this usually occurs in the vicinity of a continuum threshold and, since the resonance spontaneously appears here, can be viewed as a spectral phase transition or dynamical phase transition \cite{PPT91,TGKP16,OH17A,OH17B,DBMP08,GRHS12}.  In this picture, when the resonance is absent from the system, the system dynamics would either be oscillatory (stable) or dissipative, but non-Markovian (i.e., governed by reversible dynamics).  But when the resonance appears, one would generally expect the dominant dynamics to be Markovian, with the exponential decay width determined by the imaginary part of the complex eigenvalue.  However, it should be emphasized that, somewhat similar to the picture in the quantum phase transition, the dynamics in the {\it immediate} vicinity of the EP2A are not necessarily exponential, even after the resonance has appeared.  We will show this explicitly in what follows.  (As a brief aside, we note that in the context of parity-time (PT) symmetric systems, the EP2A is associated with PT-symmetry breaking \cite{GGH15,BenderBook}.)

Next, we denote an exceptional point at which two resonance eigenstates coalesce before forming two different resonances as an EP2B.  Because the resonance exists on either side of the EP2B, this case is not associated with the same shift from non-Markovian to Markovian dynamics as in the case of the EP2A.  However, the EP2B is associated with the phenomenon of resonance trapping \cite{Rotter_review,RotterBird,PRSB00,Past07} and qualitative features of the exponential decay can change on either side of the transition \cite{Past07}.  Further, the dynamics in the immediate vicinity of the EP2B quite generally appears as power law-exponential decay  \cite{GW64,BG65,WKH08,CM11,HeissSS,FMCW14,Hashimoto1,Hashimoto2}, as demonstrated in a microwave billiard experiment \cite{Dietz07,Dietz14}.  This occurs as a result of the double pole at the EP2B.

The generalization of this categorization scheme to higher-order EPs involving three or more coalescing states \cite{GraefeEP3,HW16} is fairly natural and we will comment on the higher-order cases of both A-type and B-type exceptional points in what follows.  

In Sec. \ref{sec:EP2B}  we look at a simple model for an EP2B in an open quantum system and demonstrate the appearance of power law-exponential dynamics on the primary timescale, which is replaced by inverse power law decay at very long timescales.  Then, in Sec. \ref{sec:EPNB} we discuss the generalization to higher-order B-type EPs and also briefly review their influence on some other physical quantities that have been studied in the literature, including scattering processes.  We then turn to the A-type exceptional point and show in Sec. \ref{sec:EP2A} an example in which an EP2A near the band edge gives rise to decay of the characteristic form $1 - C_1 t^{1/2} + C_2 t$.  A quick numerical simulation of the dynamics in the vicinity of an EP3A near the band edge is then presented in Sec. \ref{sec:EP3A}.  Finally, we summarize our main results in Sec. \ref{sec:conc} and give some comments on the literature and directions for future work, including a comment on potential experimental realization of these results.


\section{Quantum dynamics near an EP2B}
\label{sec:EP2B}

To study the dynamics near an EP2B, we introduce a simple model for a qubit coupled to a semi-infinite hopping chain, described by the Hamiltonian
\begin{equation}
  H_q
  	= 
	- V \left( \dA^\dagger \dB + \dB^\dagger \dA \right)
	- g \left( c_{1}^\dagger \dB + \dB^\dagger c_{1} \right)
	- J \sum_{j = 1}^\infty \left( c_j^\dagger c_{j+1} + c_{j+1}^\dagger c_j \right)
		.
\label{ham.B}
\end{equation}
Here, the intra-qubit coupling is given by $-V$ in the first term, while the chemical potential of the two qubit sites (with creation operators $\dA^\dagger, \dB^\dagger$) are both taken as zero.  The second site of the qubit $\dB$ is then coupled to the endpoint of the chain $c_1^\dagger$ with coupling parameter $-g$, while the hopping parameter along the chain is given by $-J$ in the third term.   We will measure energy in this system according to the unit $J = 1$.  Notice that no interaction terms appear in (\ref{ham.B}), so that we can work within the single-particle framework below.

This model was originally introduced by A. D. Dente and collaborators in Ref. \cite{DBMP08} to analyze a variety of parameter regimes in which the dynamics exceeds description by the usual Fermi golden rule.   For example, the authors show that a regime of anomalous diffusion exhibiting pure inverse power law decay can be associated with virtual states (solutions with real energy eigenvalues appearing on the second Riemann sheet) \cite{DBMP08} (see also \cite{GPSS13,Pastawski06,BCP10,GNOS19}).  However, the dynamics near the exceptional point for this model are first considered in detail in Ref. \cite{GO17}  We outline the main points of that calculation below.


\subsection{Model $H_q$: effective Hamiltonian, spectrum and exceptional points}
\label{sec:EP2B.eff}

As the first step in our analysis, we assume our boundary condition as an outgoing wave from the qubit region (the Siegert boundary condition) \cite{Siegert39,Hatano_eff}
\begin{equation}
  \psi (x) =
	\left\{ \begin{array}{ll}
		\psi_1				& \mbox{for $x = \dA$}	\\
		\psi_2				& \mbox{for $x = \dB$}   	\\
		C e^{ikx}				& \mbox{for $x \ge 1$}
	\end{array}
	\right.  
	.
\label{II.outgoing}
\end{equation}
This represents the initial condition we will later consider, in which the $\dA$ site of the qubit is initially occupied at $t=0$.
We next introduce the projection operator for the qubit sector of the model as
\beq
  P
  	= | \dA \ket \bra \dA | +  | \dB \ket \bra \dB | 
	,
\label{B.P}
\eeq
and the complementary projection operator for the chain (environment) portion of the model
\beq
  Q
  	= 1 - P
	=  \sum_{j=1}^\infty | j \ket \bra j |
	.
\label{B.Q}
\eeq
We can immediately obtain the energy eigenvalue associated with the $Q$ sector by solving the Schr\"odinger equation for any $x \ge 2$ as 
$\bra x | H | \psi \ket = E \bra x | \psi \ket$, which gives
\beq
  - \psi \left( x - 1 \right) - \psi \left( x + 1 \right)
  	= E \psi \left( x \right)
	.
\label{sch.eq.chain}
\eeq
Applying the plane wave solution $\psi (x) = C e^{ikx}$ from Eq. (\ref{II.outgoing}) in the above equation yields
\beq
  E \left( k \right)
  	= - 2 \cos k
		,
\label{cont.disp}
\eeq
after setting the energy units as $J=1$.

To obtain the discrete spectrum associated with the qubit, we next project out the $Q$ sector of the model by applying the Feshbach method \cite{Rotter_review,Hatano_eff,Feshbach_1,Feshbach_2,SR03}.
Doing so, we obtain an equivalent form of the Schr\"odinger equation written in the projected $P$ sector as
\beq
  H_\textrm{eff} (E_j) P | \psi_j \ket 
  	= E_j \left( P | \psi_j \ket \right)
\label{H.eff.schr}
\eeq
in which the effective Hamiltonian is obtained according to
\beqa
  H_\textrm{eff} (E)
    & 	= & PH_q P + PH_q Q \frac{1}{E - Q H_q Q} QH_q P
			\nonumber		\\
   &  	= & \left( \begin{array}{ccc}
		0	& - V   	\\
		- V		& - g^2 e^{i k}
	\end{array}
	\right)
	.
\label{B.H_eff}
\eeqa
We emphasize two points about the effective Hamiltonian.  First, note that $H_\textrm{eff}$ has dependence on its own eigenvalue through the lower-right entry $- g^2 e^{i k}$ and Eq. (\ref{cont.disp}).  Hence, Eq. (\ref{H.eff.schr}) is a non-linear eigenvalue problem.  Second, the factor $e^{i k}$ in this entry clearly represents the residual influence of the chain onto the qubit following the projection.  Notice this factor also renders $H_\textrm{eff}$ non-Hermitian.  Hence, we see that projecting out the environmental degrees of freedom has revealed the implicit non-Hermitian character of our model.

\begin{figure}[h]
\begin{minipage}{16pc}
\includegraphics[width=15pc]{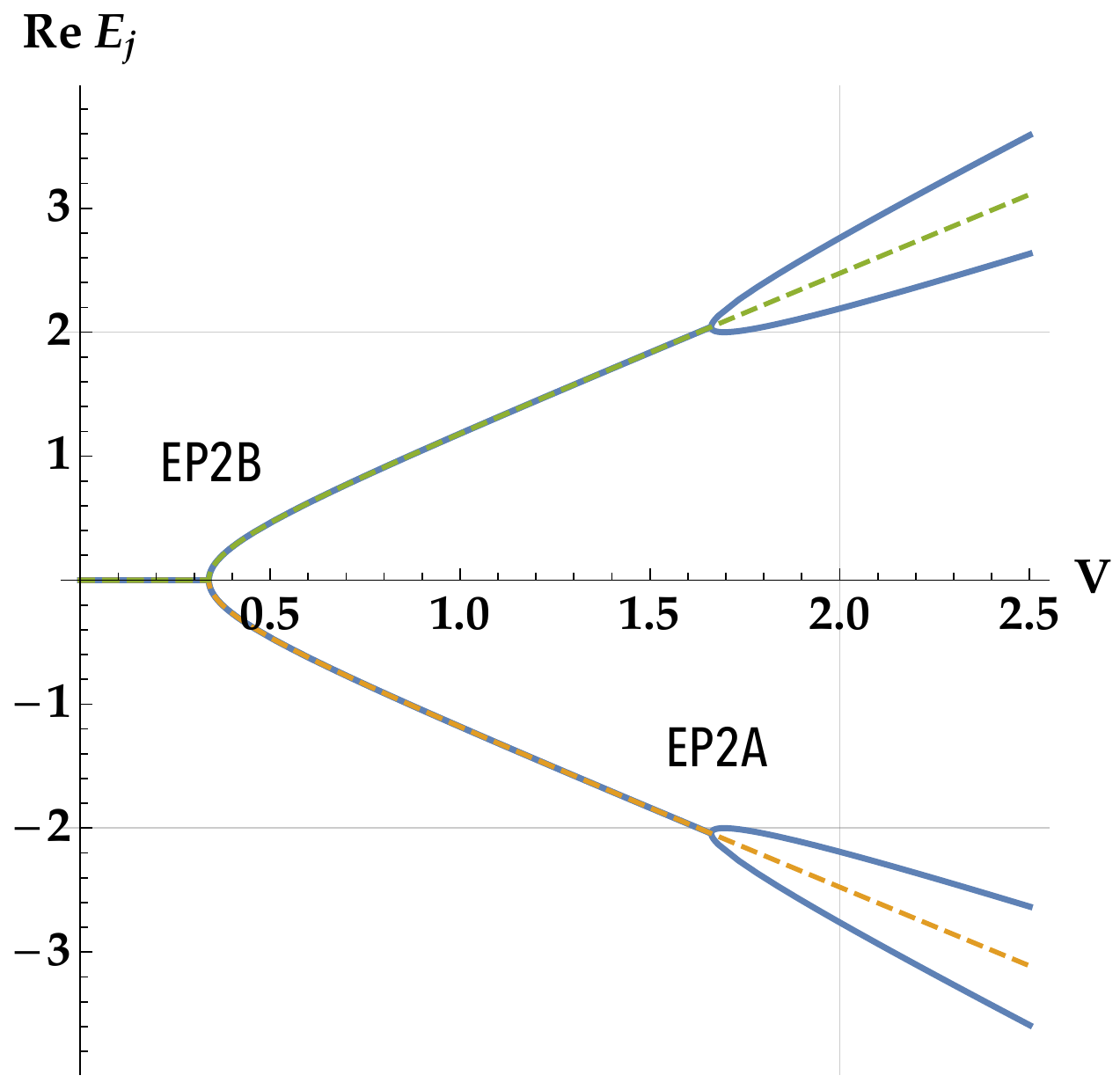}
\caption{\label{fig:B.real}
The real part of the discrete solutions for model $H_q$ from Eq. (\ref{B.E.j}) are shown as the blue, thick lines for $g=0.75$.  The location of the band edges $E = \pm 2$ are indicated by lines on both axes.  The EP2As occur at $ \bar{V}_A \approx 1.66144$ and the EP2Bs occur at $ \bar{V}_B \approx 0.33856$.  The Puiseux expansions centered on the EP2B from Eq. (\ref{B.EP.puiseux}) are shown with the dashed green and orange lines.
}
\end{minipage}\hspace{2pc}%
\begin{minipage}{16pc}
\includegraphics[width=15pc]{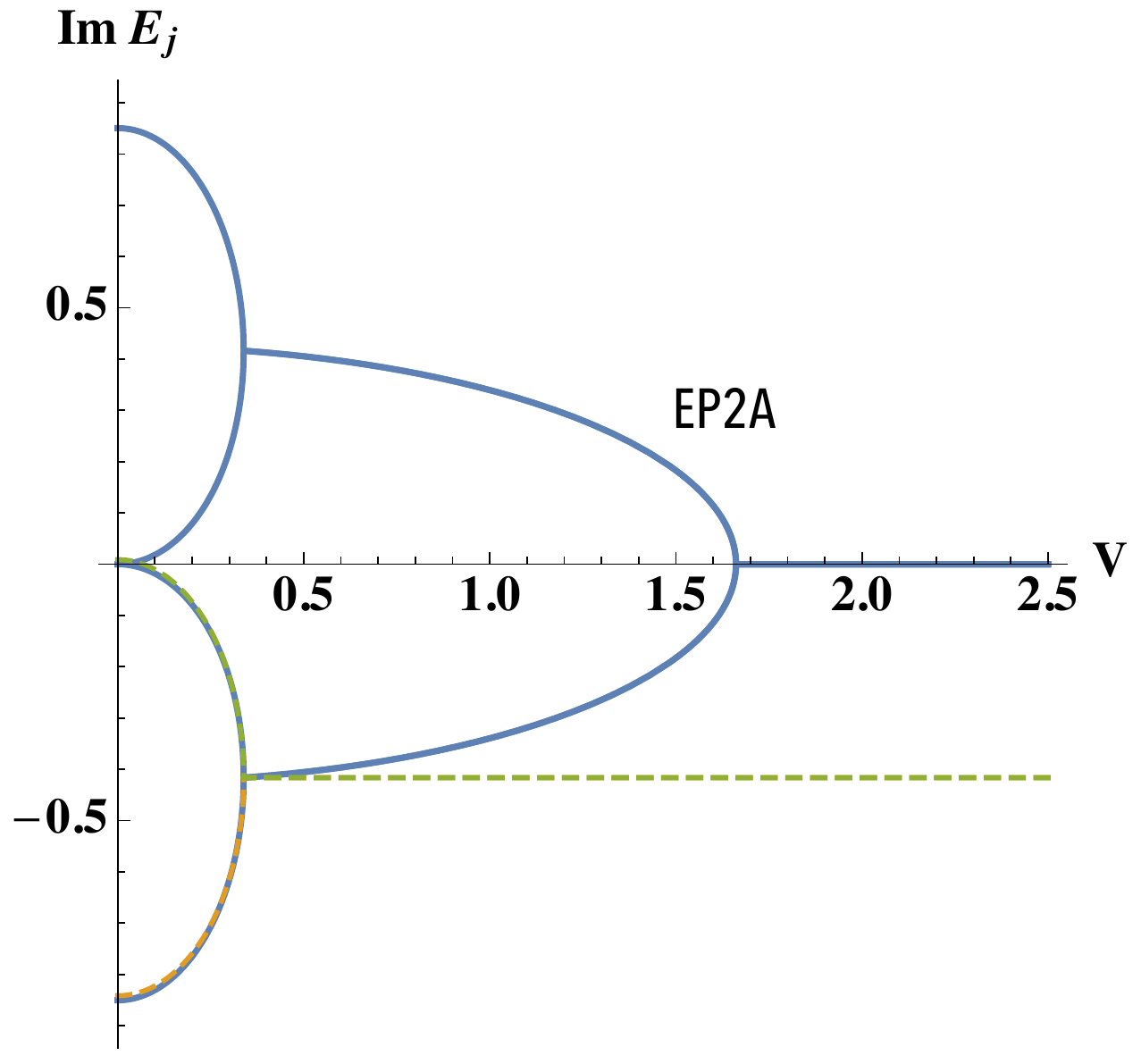}
\caption{\label{fig:B.imag}
The imaginary part of the discrete solutions for model $H_q$ from Eq. (\ref{B.E.j}) are shown as the blue, thick lines for $g=0.75$.  
The EP2As occur at $ \bar{V}_A \approx 1.66144$ and the EP2Bs occur at $ \bar{V}_B \approx 0.33856$.  The Puiseux expansions centered on the EP2B from Eq. (\ref{B.EP.puiseux}) are shown with the dashed green and orange lines.
}
\end{minipage} 
\end{figure}

Taking the determinant of Eq. (\ref{H.eff.schr}) with (\ref{B.H_eff}) and applying Eq. (\ref{cont.disp}) we obtain a polynomial equation $p_B (E_j) = 0$ for the discrete eigenvalues of $H_q$.  Here $p_B (E)$ is quartic in $E$ and given by
\begin{equation}
  p_B (E)
  	= \left( 1 - g^2 \right) E^4
		+ \left[ g^4 + \left( g^2 - 2 \right) V^2 \right] E^2 + V^4
	.
\label{B.disp.E}
\end{equation}
As the quartic takes double quadratic form, we can immediately solve  $p_B (E_j) = 0$ for
the four discrete eigenvalues, given by all four sign combinations in
\begin{equation}
  E_j
  	= \pm \sqrt{
		\frac{ \left( 2 - g^2 \right) V^2 - g^4 \pm g^2 \sqrt{g^4 + 2 \left( g^2 - 2 \right)V^2 + V^4}}
			{2 \left( 1 - g^2 \right)} }
	.
\label{B.E.j}
\end{equation}
The real and imaginary parts of these solutions are plotted as the full blue lines for the case $g=0.75$ in Figs. \ref{fig:B.real} and \ref{fig:B.imag}, respectively.
From Eq. (\ref{B.E.j}) we can now easily find the location of the exceptional points in the model.  Let $g < 1$ be fixed while $V$ is a controllable parameter.   Then there is an EP2A that appears at $V = \bar{V}_A$ with
\beq
   \bar{V}_A = 1 + \sqrt{1-g^2}
 	,
\label{B.VA}
\eeq
which appears at about $ \bar{V}_A \approx 1.66144$ in Figs. \ref{fig:B.real} and \ref{fig:B.imag}.
However, since it easier to analyze the properties near the EP2A in the simpler model studied later in Sec. \ref{sec:EP2A}, we will not focus on this here.  Instead, we primarily devote our attention to the EP2B located at
\begin{equation}
  \bar{V}_B
  	= 1 - \sqrt{1-g^2}
	,
\label{B.VB}
\end{equation}
which corresponds to the coalesced eigenvalues
\begin{equation}
  \mp \bar{E}_B
  	= \mp i \sqrt{\frac{2-g^2}{\sqrt{1-g^2}} - 2}
	 \equiv \mp i \frac{\bar{\Gamma}_B}{2}
		.
\label{B.EB}
\end{equation}
Here, the upper sign is for the two coalesced resonance states and the lower sign is for the coalescence of the associated anti-resonance states.  Note we have also defined the decay width for the coalesced resonance at the EP2B as $\bar{\Gamma}_B$.  These two EP2Bs can be seen at $ \bar{V}_B \approx 0.33856$ in Figs. \ref{fig:B.real} and \ref{fig:B.imag}.

As mentioned in the Introduction, even when the system is detuned from the precise location of the exceptional points, their existence still strongly influences the spectral features of the model.  This influence can be expressed in terms of the Puiseux expansion that gives an accurate approximation of the coalescing eigenvalues in the near vicinity of the EPs.  Following methods similar to those outlined in Refs. \cite{GGH15,GRHS12}, we obtain the Puiseux expansion for the two resonances in the present model near the EP2B as
\beq
  E = - i \frac{\bar{\Gamma}_B}{2} \pm \frac{g^2}{2 \sqrt{\left(1-g^2\right)\left(2 - g^2 - 2 \sqrt{1-g^2}\right)}}
  	\left(V^2 -  \bar{V}_B^2 \right)^{1/2}
		+ \mathcal{O}\left( V^2 -  \bar{V}_B^2 \right)
	.
\label{B.EP.puiseux}
\eeq
These are shown by the green and orange dashed lines in Figs. \ref{fig:B.real} and  \ref{fig:B.imag}.
Note that it is this Puiseux expansion that gives rise to enhanced parametric sensitivity near the exceptional points, as described in Refs. \cite{Wiersig14,Chen,Stone19,ChristoEP3,WiersigEP3}.


\subsection{Linearized eigenvalue problem for $H_q$}
\label{sec:EP2B.GEP}

The detailed calculation for the dynamics near the EP2B in the present model is slightly tedious but largely straightforward.  Here we just present a quick outline, emphasizing one point about the divergence of the norm that we think provides context about the properties of the exceptional points.  The details of the calculation can be found in Ref. \cite{GO17} (By comparison, in Sec. \ref{sec:EP2A} we will give a somewhat more detailed recounting of the calculation for the dynamics near the EP2A, as that situation is a bit more subtle.)

Following Ref. \cite{GO17}, as a first step in analyzing the dynamics, we transform the non-linear eigenvalue problem in Eq. (\ref{H.eff.schr}) into a generalized linear eigenvalue problem, according to the formalism of Ref. \cite{HO14}.
To do this we introduce the variable 
\beq
  \lambda \equiv e^{ik}
	,
\label{lambda}
\eeq
which is convenient to work with below.  Our objective now is to transform the original non-standard eigenvalue problem Eq. (\ref{H.eff.schr}) in terms of $E$ into an equivalent linear eigenvalue problem in $\lambda$.

The continuum eigenvalue in Eq. (\ref{cont.disp}) can be immediately rewritten in terms of $\lambda$ as the simple expression
\beq
  E = - \lambda - 1/\lambda
\label{cont.disp.lambda}
\eeq
Applying this in Eq. (\ref{H.eff.schr}), the eigenvalue equation for the effective Hamiltonian can be shown to be a $2 \times 2$ quadratic eigenvalue problem in $\lambda$.  Following the standard method to linearize the quadratic eigenvalue problem \cite{TM01,SM_book}, this in turn can be transformed into a $4 \times 4$ generalized linear eigenvalue problem of the form 
\beq
  \left( F - \lambda G \right) | \Psi \ket 
  	= 0
\label{GEP}
\eeq
in which
\beq
  F = 
	\left[ 
	\begin{array}{cccc}
		0	& 0	& 1		& 0				\\
		0	& 0	& 0		& 1				\\
		1 	& 0	& 0		& -V				\\
		0	& 1	& -V		& 0
			\end{array}
		\right]
			\ \ \ \ \ \ \ \ \ \ \ \ \ \ \ \ \ \ \ 
  G = 
	\left[ 
	\begin{array}{cccc}
		1	& 0	& 0		& 0				\\
		0	& 1	& 0		& 0				\\
		0 	& 0	& -1		& 0				\\
		0	& 0	& 0		& g^2 - 1
			\end{array}
		\right]
	.
\label{B.A.B}
\eeq
and
\beq
  | \Psi \ket
	= 	\left[ 
	\begin{array}{c}
		\bra d_1 | \psi \ket	\\
		\bra d_2 | \psi \ket	\\
		\lambda \bra d_1 | \psi \ket 	\\
		\lambda \bra d_2 | \psi \ket
			\end{array}
		\right]
	.
\label{B.Psi.defn}
\eeq
Notice the key point that the dimension of the eigenvalue problem in Eq. (\ref{GEP}) is four, which is in one-to-one correspondence with the number of discrete eigenvalues of the original Hamiltonian $H_q$.  We also emphasize that although the $F$ and $G$ matrices are Hermitian, the {\it generalized} eigenvalue problem in Eq. (\ref{GEP}) is non-Hermitian because $G$ is not the identity matrix.  To see this another way, one could rewrite Eq. (\ref{GEP}) in the equivalent form $ \left( G^{-1} F - \lambda I_4 \right) | \Psi \ket = 0$, in which $G^{-1} F$ is explicitly non-Hermitian.

Setting the determinant of Eq. (\ref{GEP}) equal to zero we obtain the four $\lambda$ eigenvalues as
\begin{equation}
  \lambda_{j}
  	= \pm \sqrt{
		\frac{ V^2 + g^2 - 2 \pm \sqrt{g^4 + 2 \left( g^2 - 2 \right)V^2 + V^4}}
			{2 \left( 1 - g^2 \right)} }
	.
\label{B.lambda.j}
\end{equation}
Plugging these into Eq. (\ref{cont.disp.lambda}) verifies that they are equivalent to the original $E$ eigenvalues in Eq. (\ref{B.E.j}).  
We can also now obtain the coalesced eigenvalues $\pm \bar{\lambda}_B $ at the EP2B as well as a Puiseux expansion, similar to Eq. (\ref{B.EP.puiseux}); we leave the details to Ref. \cite{GO17}.



\subsection{Dynamics near the EP2B: intermediate timescale}
\label{sec:EP2B.surv}

We now evaluate the influence of the EP2B on the dynamics in the qubit sector of the model $H_q$.
To simplify our analysis we choose as an initial state $| \dA \ket$; although we emphasize the results would be qualitatively similar for the $| \dB \ket$ state or any linear combination within the qubit sector. 
The survival probability for this state is given by $P(t) = | A(t) |^2$ in which
\beq
  A(t) = \bra \dA | e^{-i H_q t} | \dA \ket
  	.
\label{B.surv}
\eeq

For very short evolutions, it is known that quantum systems universally exhibit parabolic dynamics; the exceptional point has no particular influence on the dynamics on this timescale.  For the present model, Eq. (\ref{B.surv}) can easily be evaluated to show that $P(t) \approx 1 - V^2 t^2$ for $t \ll 1/\sqrt{2} V$.
This short-lived effect has been detected in a handful of experiments \cite{short_time_expt,CrespiExpt} and
gives rise to the so-called quantum Zeno effect, with inhibited decay upon repeated, consecutive measurements \cite{SudarshanZeno,SudarshanZeno2,KK96,SegalZeno,LonghiZenoExpt,MurchZeno}.

Let us turn now to the intermediate timescale dynamics, on which the influence of the EP2B is directly pronounced.
Following the formalism of Ref. \cite{HO14}, in the context of the generalized eigenvalue problem Eq. (\ref{GEP}),
we can write the survival amplitude as
\begin{eqnarray}
  A(t) 
     &	= & \bra \dA | e^{-i H_q t} | \dA \ket
     						\nonumber  \\
     &	= & \frac{1}{2 \pi i} \sum_{j = \{ s, \pm \}} 
		\int_\mathcal{C} d \lambda \left( - \lambda + \frac{1}{\lambda} \right) 
			\exp \left[ i \left( \lambda + \frac{1}{\lambda} \right) t \right]
				\bra \dA | \psi_j \ket \frac{\lambda_j}{\lambda - \lambda_j} \bra \tilde{\psi}_j | \dA \ket
	,
\label{B.GEP.surv}
\end{eqnarray}
which is equivalent to the usual integral over the Green's function, but here the integration is rewritten in terms of the $\lambda$ variable and further decomposed into contributions from each of the four $| \psi_j \ket$ eigenstates.  We note the presence of the norm of the four eigenstates $ \bra \tilde{\psi}_j | \dA \ket \bra \dA | \psi_j \ket$ in this expression, in which $\bra \tilde{\Psi}_j |$ denotes the left eigenstate for the corresponding left eigenvalue equation 
$\bra \tilde{\Psi} | \left( F - \lambda G \right) = 0$.

The norm can be explicitly obtained from the condition $\bra \tilde{\Psi}_j | G | \Psi_j \ket = 1$ and the eigenvalue equation (\ref{GEP}) as
\begin{equation}
  \bra \tilde{\psi}_j | \dA \ket \bra \dA | \psi_j \ket =
  \bra \dA | \psi_j \ket^2
  	= \frac{V^2 \lambda_j^2}
		{1 + \left( 1 + g^2 + V^2 \right) \lambda_j^2 
			- \left( 1 - 2g^2 + V^2 \right) \lambda_j^4 - \left( 1 - g^2 \right) \lambda_j^6}
		.
\label{B.GEP.eigenket.dA}
\end{equation}
Similar to the Puiseux expansion for $E_j$, 
we can reparameterize the $V$ parameter near the EP2B as
$  V = \bar{V}_B + \delta $
and then expand the eigenstate norm near the EP2B as
\begin{eqnarray}
	\bra \dA | \psi_{\sigma,\pm} \ket^2
  	= \frac{1}{4} \left[ \mp \frac{1}{\sqrt{2 \delta}} \sqrt{1 - g^2 - \sqrt{1-g^2}} + 1 
		+ O(\delta^{1/2})
			\right]
		.
\label{B.GEP.eigenket.dA.exp}
\end{eqnarray}
Note this expression is divergent as $\delta \to 0$, which is a general property of the EP.  However, also note that the sign of the divergent term is opposite for the two lines of this equation, corresponding to the two coalescing states. Indeed, in physical quantities, the divergent contributions must always cancel at the EP.  

\begin{figure}[h]
\begin{minipage}{16pc}
\includegraphics[width=16pc]{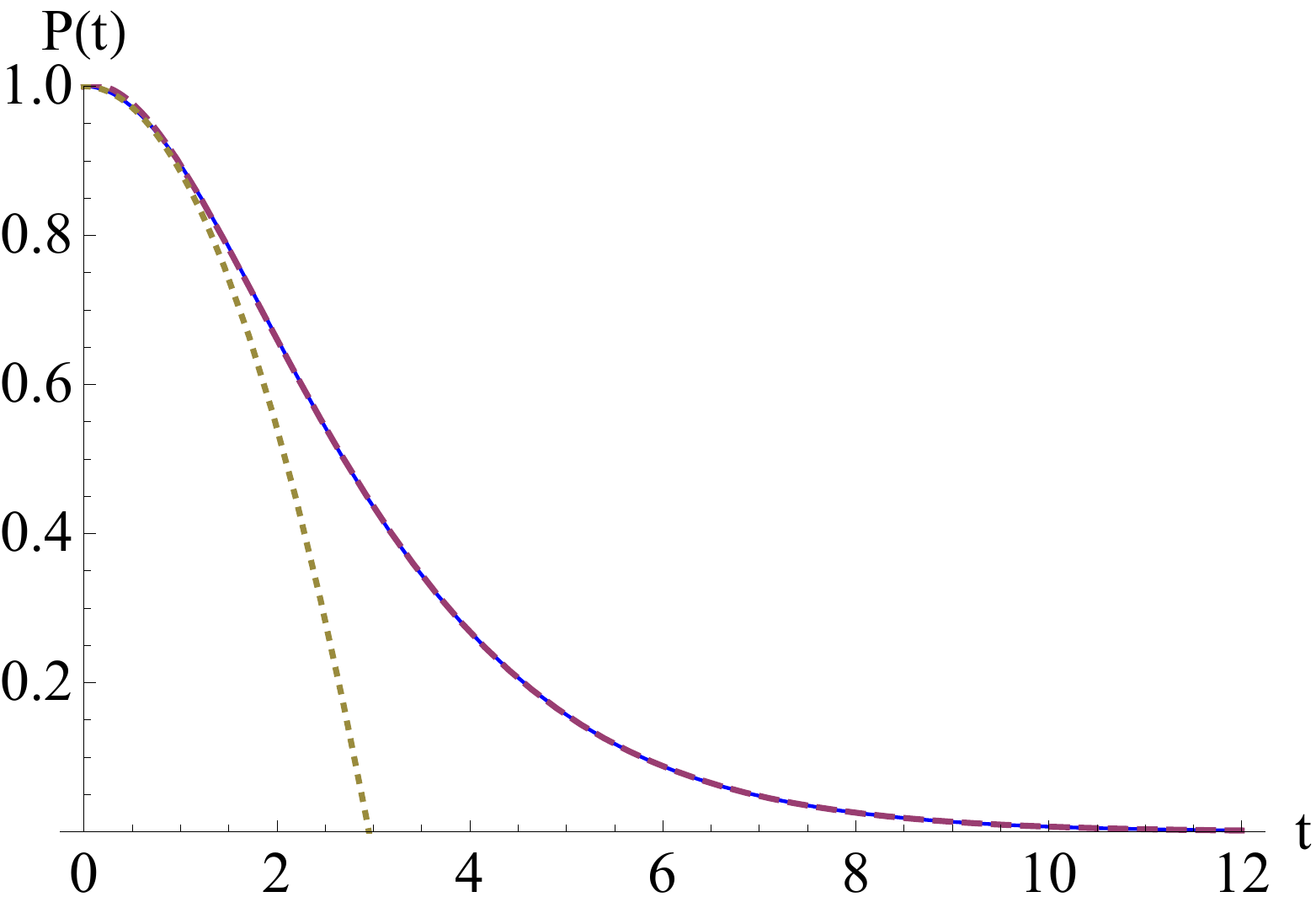}
\caption{\label{fig:EP2B.1}
Survival probability $P(t)$ for $g=0.75$ and $V=0.3385622$, near the EP2B at $\bar{V}_B \approx 0.3385620$ .  The purple dashed curve is the approximation from Eq. (\ref{B.surv.EP2B}), which agrees well with the numerical integration (solid blue curve).  The beige dotted curve shows the parabolic early-time dynamics.}
\end{minipage}\hspace{2pc}%
\begin{minipage}{16pc}
\includegraphics[width=16pc]{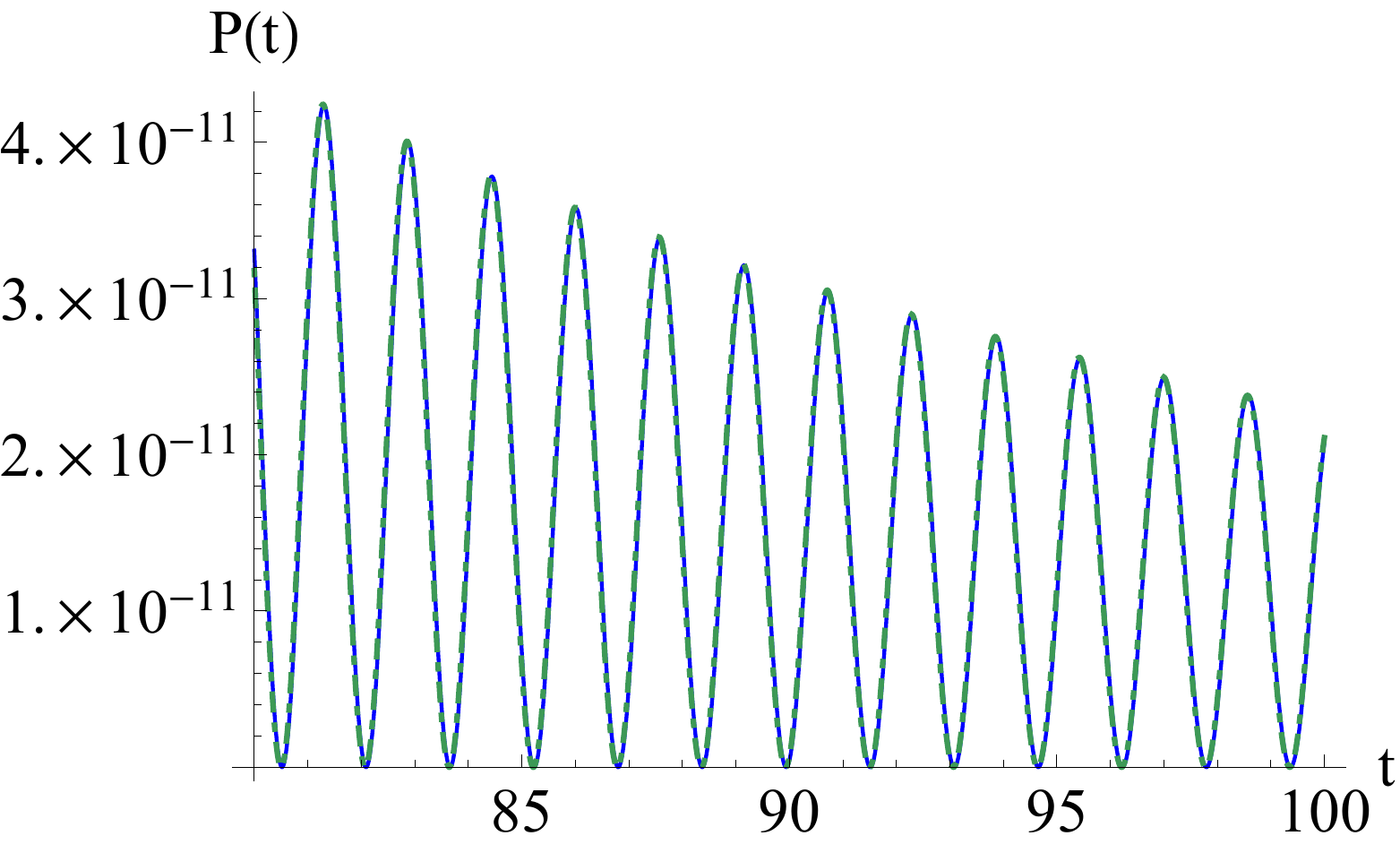}
\caption{\label{fig:EP2B.2}
Long-time survival probability $P(t)$ exactly at the EP2B $\bar{V}_B \approx 0.3385620$ for $g=0.75$.
The approximation corresponding to Eq. (\ref{B.surv.long}) is shown as the green dash-dotted curve and again agrees well the numerical integration, shown as the solid blue curve.
}
\end{minipage} 
\end{figure}

Applying the expansion Eq. (\ref{B.GEP.eigenket.dA.exp}) as well as the Puiseux expansion for the $\lambda_j$ eigenvalues, the summed terms under the integral in Eq. (\ref{B.GEP.surv}) can in turn be expanded as
\begin{eqnarray}
  \sum_{j = \{ s, \pm \}} 
	\frac{\lambda_j}{\lambda - \lambda_j} \bra \dA | \psi_j \ket^2
	     = \frac{\lamB^4 + \lambda^2}{\left( \lamB^2 - \lambda^2 \right)^2} + O (\delta)
		,
\label{B.GEP.surv.factor}
\end{eqnarray}
in which the double-pole is the direct result of the EP2B.  Now evaluating the residue for the double pole, we find the survival probability near the EP2B takes the characteristic form
\begin{eqnarray}
  P(t) 
	 \left[ 1 + D_1 t + D_2 t^2  \right]
     		e^{- \bar{\Gamma} t }
	.
\label{B.surv.EP2B}
\end{eqnarray}
in which
\beq
  D_1 = \frac{g^2 \left( 1 + \sqrt{1 - g^2} \right)}{2 \left( 1 - g^2 \right)^{3/4}}
  			\ \ \ \ \ \ \ \ \ \ \ \ \ \ \ \ \ \ \ \ 
  D_2 = \frac{g^4 \left( 2 -g^2 + 2 \sqrt{1 - g^2} \right)}{16 \left( 1 - g^2 \right)^{3/2}}
  	.
\eeq
We show this prediction for the dynamics matches very well with a numerical integration for the case $g=0.75$
in Fig. \ref{fig:EP2B.1}.

This result for the dynamics near an EP2B has been confirmed in the microwave cavity experiment in Refs. \cite{Dietz07,Dietz14}.

\subsection{Dynamics near the EP2B: asymptotic timescale}
\label{sec:EP2B.long}

Similar to the short-time parabolic dynamics mentioned earlier, it can be shown that quantum systems quite universally exhibit non-exponential decay on very large timescales \cite{GPSS13,Khalfin,Fonda,Muga_review,Hack,GCMM95,Cavalcanti,MDS95}  However, under most circumstances this effect would be extremely difficult to detect because it does not appear until after many lifetimes of the exponential decay have passed, by which time almost nothing of the initial state remains to be measured in experiment \cite{Muga_review}

For the present model, it is shown in Ref. \cite{GO17} that at the EP2B the long-time dynamics exhibit the behavior
\beq
  P(t) \sim t^{-3} \cos^2\left(2t + \pi/4\right) 
\label{B.surv.long}
\eeq
as shown for the case $g = 0.75$ in Fig. \ref{fig:EP2B.2}.  We note here that the inverse power law decay $1/t^3$ is rather typical for long-time dynamics in 1-D systems \cite{SZMG17,GPSS13,GCMM95,MDS95} and the oscillations in Eq. (\ref{B.surv.long}) emerge here due to the equal influence from the two band edge contributions at $E = \pm 2$.
Hence, unlike the intermediate time dynamics, the EP2B doesn't seem to have any pronounced influence on the long-time dynamics; at least, not in the present circumstances.  However, for the case of the EP2A in Sec. \ref{sec:EP2A}, we will find the long-time dynamics can be significantly enhanced compared to the usual case.


\section{Intermediate Discussion: Dynamics, scattering and other signatures of EPNBs}
\label{sec:EPNB}

In Sec. \ref{sec:EP2B}, we saw that near the EP2B the usual exponential dynamics were replaced with 
power law-exponential decay on the primary timescale during which most of the dissipation occurs.  While from an engineering perspective we can view this as a potential tool for designing microscopic systems with desired dynamical properties, from a physicist's perspective we could view the power-law-exponential dynamics in Eq. (\ref{B.surv.EP2B}) as a means to identify (prove) the presence of an EP2B.   Further, it is immediately clear how this result should generalize for the case of $N$ coalescing resonances at an EPNB.  This would lead to an $N$th order pole in the integrand for the survival amplitude, yielding a dynamical signature of the form $(1 + C_1 t + \dots + C_{2N-2} t^{2(N-1)}) e^{- \Gamma t}$ in the resulting survival probability \cite{GW64}.
Hence, in principle, the existence of the EP and its order could both be determined from the power law-exponential decay, without the need to tune the parameters to the exact location of the EPNB, which might be challenging.

A few other methods have been proposed to determine the presence of 
an EP2B without needing to tune to its precise location.  The first would be to dynamically encircle the EP in the parameter space, and observe the resulting modification of the usual adiabatic evolution scheme \cite{GMM13,Rotter15,HeissNat,WunnerEncircle,RotterNat,XuNat,Liu21}.  In this case, it can be shown that the final state at the end of the adiabatic evolution depends only on the direction of encirclement, not the initial state one populates at the beginning of the cycle.  This has been confirmed in several laboratory experiments \cite{RotterNat,XuNat,Liu21}.  See also the discussion in App. A of \cite{GO17}.

A second method to infer the existence of a nearby EP2B would be to measure its signature influence on the scattering cross section.  As first reported by E. Hern\'andez, et al, in Ref \cite{HJM00} in the context of scattering from a spherical double barrier potential, a double pole in the scattering matrix results in a cross section with a characteristic split-peak profile.  As argued by W. D. Heiss and G. Wunner in Ref.\cite{HW14}, this split-peak can be viewed in relation to the well-known asymmetric Fano-Friedrichs profile, which appears due to quantum interference between two different states interacting with the same decay channel, or between two different decay paths for the same state into a single channel.  Finally, A. Ben-Asher, et al, propose to observe the split-peak profile in resonance tunnelling experiments by inducing an EP2B through laser action in diodes and similar systems \cite{BSUSM20}.

It has also been shown that B-type exceptional points can induce significant enhancement of spontaneous emission in their vicinity \cite{Lin16,Pick17}.  Related to this, it is recently shown in Ref. \cite{Dunham} that 
an EP ring appears in the spectrum of a magnetically-sensitive doping impurity in a polyacetylene molecule, which results in a giant response in the single-spin electron resonance as measured by STM probe.  This result could potentially enable single-spin detection in such experiments  \cite{Dunham}.


\section{Quantum dynamics near an EP2A}
\label{sec:EP2A}

To study the dynamics near an EP2A, we now turn to a model in which a simple quantum emitter (or quantum dot) is coupled to the endpoint of a semi-infinite chain.  Our Hamiltonian in this case is given by
\beq
  H_d
  	= \epsd d^\dagger d - J \sum_{j=1}^{\infty} \left( c_j^\dagger c_{j+1} + c_{j+1}^\dagger c_j \right)
		- g \left( c_1^\dagger d + d^\dagger c_1 \right)
	.
\label{ham.A}
\eeq
Here the quantum dot with variable potential $\epsd$ is described with the creation operator $d^\dagger$ and is coupled to the endpoint of the chain with coupling strength $g$ in the third term.  The hopping strength along the semi-infinite chain is given by $J$ in the second term, just the same as the previous model studied in Sec. \ref{sec:EP2B}.  Again, we measure energy in units of $J=1$.


\subsection{Eigenvalues and effective Hamiltonian}
\label{sec:EP2A.Heff}

Similar to our initial steps for the previous model, we first write our outgoing wave boundary condition for model $H_d$ as
\beq
  \psi (x) \equiv
  	\bra x | \psi \ket
	=  \left\{ \begin{array}{ll}
		\psi_d					& \mbox{for $x = d$}   	\\
		C e^{ikx}					& \mbox{for $x \ge 1$}
	\end{array}
	\right.  
	.
\label{A.outgoing}
\eeq
Evaluating the Schr\"odinger equation along the chain for any site $x \ge 2$ yields the same continuum dispersion equation $E(k) = - 2 \cos k $ as we obtained for the previous model in Eq. (\ref{cont.disp}).

In the next step, we write the projection operators for the present model.  The projection into the discrete sector associated with the quantum dot is written simply
\beq
  P
  	= | d \ket \bra d |
\label{A.P}
\eeq
while again the projection associated with the continuum is given by
\beq
  Q
  	= 1 - P
	=  \sum_{j=1}^\infty | j \ket \bra j |
	.
\label{A.Q}
\eeq
We can again obtain an effective Hamiltonian in the $Q$ subspace after projecting out the $P$ sector according to the Feshbach method by writing
\beqa
  H_\textrm{eff} (E)
    & 	= & P H_d P + P H_d Q \frac{1}{E - Q H_d Q} Q H_d P
			\nonumber		\\
   &  	= & \epsd - g^2 e^{ik}
	.
\label{A.H_eff}
\eeqa
Applying this in the Schr\"odinger equation (\ref{H.eff.schr}) and applying Eq. (\ref{cont.disp}), we obtain the polynomial dispersion equation, which is quadratic  for the present model.  The two resulting eigenvalues are given by
\beq
  E_{\pm}
  	= \frac{\epsd \left( 2 - g^2 \right) \pm g^2 \sqrt{\epsd^2 - 4 \left( 1 - g^2 \right)} }
			{2 \left( 1 - g^2 \right)}
	.
\label{A.E.pm}
\eeq
Along with these, we can write the associated wave vectors $k_\pm$, which are obtained through $E_\pm = - 2 \cos k_\pm$.  

The location of the exceptional points can be immediately obtained from Eq. (\ref{A.E.pm}) as
\beq
  \epsd = \pm 2 \sqrt{1 - g^2},
\label{A.EPs}
\eeq
in which we assume that $\epsd$ is an experimentally accessible parameter while $g$ is fixed.  We immediately observe that for $g < 1$, these EPs are real-valued, while for $g > 1$ they escape into the complex plane (the parameter choice $g=1$ gives a special case for which the quadratic dispersion reduces to a linear polynomial).  For our purposes, we will assume the condition $g < 1$ is always satisfied.  Under this assumption, both exceptional points in Eq. (\ref{A.EPs}) denote EP2As at which two virtual states coalesce before forming a resonance/anti-resonance pair, or vice-versa.

In what follows, we will evaluate the characteristic dynamics near the exceptional point, focusing particularly on the case $g \ll 1$ for which the EP2As in turn lie near their respective thresholds $\epsd \approx \pm 2$.  In this case, as shown in Ref. \cite{GO17} both the exceptional point itself and the threshold play key roles in shaping the dynamics.  For definiteness, from this point we will focus on the lower exceptional point, which we denote $\epsd = \epsEP$ with
\beq
  \epsEP
  	\equiv - 2 \sqrt{1 - g^2}
	.
\label{eps.EP}
\eeq
(Of course, the dynamics near the other EP2A would be similar.)
In the case $\epsd = \epsEP$, the two eigenvalues coincide at
\beq
  E_+ (\epsEP) = E_- (\epsEP) =
  \eEP
  	\equiv - \frac{2 - g^2}{\sqrt{1 - g^2}}
	.
\label{E.EP}
\eeq


\subsection{Linearized eigenvalue problem for $H_d$}
\label{sec:EP2A.GEP}

From this point, if we were to attempt to evaluate the dynamics near the EP2A just applying the usual Green's function technique, we would encounter difficulties due to the exceptional point appearing near the branch cut.  To circumvent these, we will analyze the problem in terms of the linearized eigenvalue problem similar to our work on the EP2B problem in Sec. \ref{sec:EP2B}.  The fact that the contributions from the two eigenstates come pre-decomposed in the integrand enables us to rewrite the integration so that we can deal with the double pole and the branch point at separate steps in the analysis.

Just as we did in Sec. \ref{sec:EP2B}, we follow Ref. \cite{GO17} by introducing the variable 
$  \lambda \equiv e^{ik} $.
As before, the continuum eigenvalue in Eq. (\ref{cont.disp}) can then be immediately rewritten as
\beq
  E = - \lambda - 1/\lambda
  	.
\eeq
The Schr\"odinger equation (\ref{H.eff.schr}) with the effective Hamiltonian (\ref{A.H_eff}) corresponding to $H_d$ can be rewritten into a quadratic eigenvalue problem in terms of $\lambda$ in the form
\beq
  \left[ \left( 1 - g^2 \right) \lambda^2 + \epsd \lambda  + 1 \right] P | \psi \ket
  	= 0
	.
\label{A.GEP.eq}
\eeq
Again following the standard method to linearize the quadratic eigenvalue problem \cite{SM_book}, 
we rewrite Eq. (\ref{A.GEP.eq}) in the form
\beq
  \left( F - \lambda G \right) | \Psi \ket 
  	= 0
\label{A.GEP}
\eeq
in which the $2 \times 2$ $F$ and $G$ matrices are given by
\beq
  F
  	= \left[ \begin{array}{cc}
		0  	&  1		\\
		1  	&  \epsd
		\end{array} \right]
				\ \ \ \ \ \ \ \ \ \ \ \ \ \ \ \ \ \ \ \ 
  G
  	= \left[ \begin{array}{cc}
		1  	&  0		\\
		0  	&  -1 + g^2
		\end{array} \right]
\label{I.A.B}
\eeq
and the generalized eigenstate $ | \Psi \ket$ takes the form
\beq
  | \Psi \ket
  	\equiv \left[  \begin{array}{c}
		\bra d | \psi \ket	\\
		\lambda \bra d | \psi \ket
		\end{array} \right]
	.
\label{Psi.defn}
\eeq
Setting the determinant of (\ref{A.GEP}) to zero yields the $\lambda$ eigenvalues as
\beq
  \lambda_\pm
  	= \frac{- \epsd \mp \sqrt{\epsd^2 - 4 \left( 1 - g^2 \right)} }{2 \left( 1 - g^2 \right)}
\label{A.lambda.pm}
\eeq
Finally, we can obtain the norm of the generalized eigenstates $| \Psi \ket$ from the natural orthonormalization condition
\begin{equation}
  \bra \tilde{\Psi}_i \left| G \right| \Psi_j \ket = \delta_{i,j}
  	,
\label{B.norm}
\end{equation}
in which $\bra \tilde{\Psi}_i |$ is again the left-eigenstate obtained from the left-eigenvalue problem $\bra \tilde{\Psi} | \left( F - \lambda G \right) = 0$.  It is not difficult to show that
$\bra \tilde{\Psi} | = | \Psi \ket^T$ \cite{HO14,GO17}.
Evaluating the norm from Eq. (\ref{B.norm}), we find
\beq
  \bra d|\psi_\pm\ket\bra{\tilde\psi_\pm}|d\ket =  \bra d|\psi_\pm\ket^2
	  = \frac{1}{1 - \left( 1 - g^2 \right) \lambda_\pm^2}
	.
\label{A.norm}
\eeq


\subsection{Exceptional point properties in the $\lambda$ notation}
\label{sec:EP2A.puiseux}



Here we derive a few expressions needed for the survival probability calculation.

Just as the energy eigenvalues coincided at the exceptional point $\epsd = \epsEP$, the two $\lambda$ eigenvalues also coincide, giving
\beq
  \lambda_+ (\epsEP)  = \lambda_- (\epsEP) =
  \lamEP
  	\equiv - \frac{2}{\epsEP}
	= \frac{1}{\sqrt{1 - g^2}}
	.
\label{lambda.EP}
\eeq
Next, to evaluate the dynamics near the EP we parameterize the $\epsd$ variable by writing
\beq \label{eq:ExpEP}
  \epsd = \epsEP + \delta
\eeq
in which $\delta$ is taken to be a small, externally controllable parameter.  
We then apply this reparameterization and expand Eq. (\ref{A.lambda.pm}) to obtain the Puiseux series for the $\lambda$ eigenvalues as
 \beq \label{eq:lamrd}
\lambda_\pm  = \lamEP\left(1\pm i\delta^{1/2}\lamEP^{1/2}- \frac{\lamEP\delta}{2}\right) + O(\delta^{3/2})
	.
\eeq
Similarly, we expand the norm from Eq. (\ref{A.norm}) to find
\beq
  \bra d | \psi_\pm \ket^2 
	= \frac{1}{2} \left( 1\pm \frac{i}{ (\lamEP\delta)^{1/2} }\right) + O(\delta^{1/2})
	.
\label{A.norm.exp}
\eeq
We emphasize that when we calculate the survival probability below, the divergent contributions in this expression from the two states will cancel.


\subsection{Pure non-Markovian dynamics near the EP2A: threshold influence}
\label{sec:EP2A.dynamics}

Now we move on to the actual evaluation of the survival probability, 
which is again written $P(t) = | A(t) |^2$, in which
\beq
A(t)
	= \bra d | e^{-i H_d t} | d \ket
	.
\label{A.surv.amp}
\eeq
is the survival amplitude of the initially-prepared $| d \ket$ state.  

Similar to our analysis in Sec. \ref{sec:EP2B}, we can easily perform an operator expansion of the exponential in Eq. (\ref{A.surv.amp}) to find that for early times, the system follows the usual parabolic decay of the form
\beq
P_Z (t)
	\approx 1 - g^2 t^2
	,
\label{A.zeno}
\eeq
for $t \lesssim T_Z$, in which
\beq
  T_Z
  	= \frac{1}{\epsd}
	.
\label{T.Z}
\eeq
Again, this particular result for the early timescale dynamics has no particular relationship to the EP2A and it would be difficult to detect in many circumstances.  However, we will discover a situation below in which the Zeno timescale dynamics are significantly enhanced near the EP2A, although that situation is not our primary focus here.





\begin{figure}[h]
\includegraphics[width=20pc]{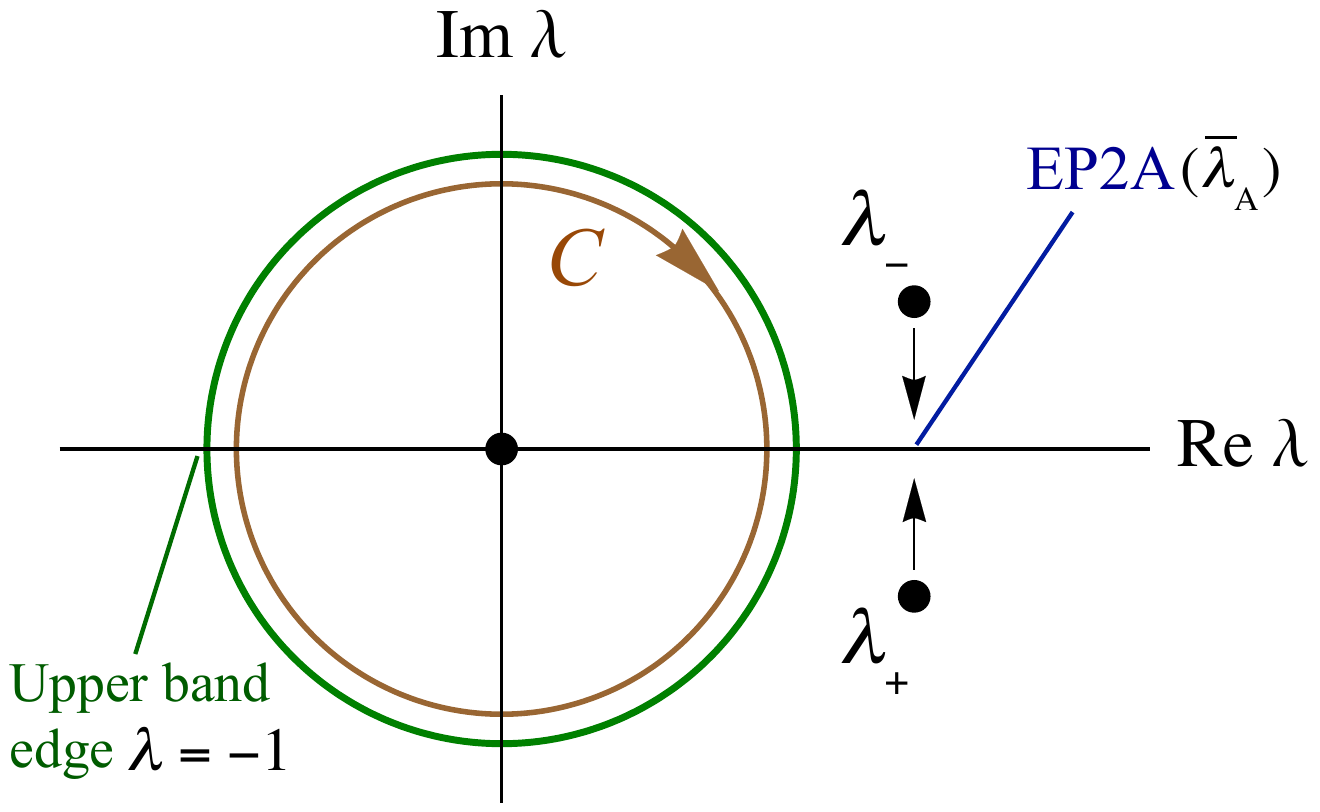}\hspace{2pc}%
\begin{minipage}[b]{14pc}
\caption{\label{fig:cont.lambda.1}Integration contour $C$ in the complex $\lambda$ plane.  The continuum coincides with the unit circle, indicated in green. The lower (upper) band edge has been mapped from $E=-2$ ($E=2$) to $\lambda = 1$ ($\lambda = -1$).  An essential singularity appears at the origin, while another occurs at infinity.
The $\lambda_\pm$ eigenvalues approach the coalescent eigenvalue $\lamEP$ on the real axis as $\epsd \rightarrow \epsEP$.}
\end{minipage}
\end{figure}

Similar to our development in Sec. \ref{sec:EP2B}, to evaluate the influence of the EP2A on the dynamics on the intermediate timescale we find it useful to write the survival amplitude as a sum of the individual contributions from the two discrete eigenstates according to the method developed in Ref. \cite{HO14} .  Following this method as originally worked out in Ref.  \cite{GO17} , we can write the survival amplitude for the initial state $| d \ket$ as
\beqa
  A(t) & \equiv & \bra d|e^{-i H_I t}|d\ket   \nonumber \\
  & 	= &\frac{1}{2\pi i}
	\sum_{j=\pm} \int_{C}d\lambda \,\left(-\lambda+\frac{1}{\lambda}\right)\exp\left[i\left(\lambda+\frac{1}			{\lambda}\right)t\right]
	\bra d|\psi_j\ket\frac{\lambda_j}{\lambda-\lambda_j}\bra\tilde{\psi}_j |d\ket.
\label{A.surv}
\eeqa
in which the contour $C$ in the complex $\lambda$ plane is indicated by the brown, clockwise-oriented circle just inside the unit circle in Fig. \ref{fig:cont.lambda.1}.  
Further, the summation over $j$ indicates the two discrete eigenstates of the Hamiltonian; these consist of two virtual bound states when the condition $\epsd \lesssim \epsEP$ holds, or a resonance/anti-resonance pair when $\epsd \gtrsim \epsEP$ holds.  The dynamics are qualitatively similar in these two cases very near to the exceptional point.

Applying the expansions Eq. (\ref{eq:lamrd}) [for $\lambda_j$] and Eq. (\ref{A.norm.exp}) [for the norm] to the integrand in Eq. (\ref{A.surv}), we can expand the survival amplitude in terms of the small parameter $\delta$ to find
\beq
  A(t) \approx \frac{1}{2\pi i} 
  	\int_{C}d\lambda \,\left(-\lambda+\frac{1}{\lambda}\right)
		\exp\left[i\left(\lam+\frac{1}{\lam}\right)t\right] \frac{-\lamEP^2}{(\lam-\lamEP)^2} +O(\delta^{1/2})
	,
\label{A.surv.exp}
\eeq
in which we remind the reader that $\delta$ is defined according to our reparameterization 
$\epsd = \epsEP + \delta$ from Eq. (\ref{eq:ExpEP}).  Notice in the first term of the above result that the divergent terms from the norm of the eigenstates have cancelled; further, the lowest-order term is independent of $\delta$.

At this point, if the EP2A wasn't close to the band edge, we could just follow the usual analysis and deform the contour to take the residue at the double pole, similar to our calculation for the EP2B in Sec. \ref{sec:EP2B}.
However, as shown in Ref. \cite{GO17}, not only is the result of such a calculation inaccurate, it is actually non-unitary.   This is because, in the present situation, the influence on the dynamics from the continuum threshold (branch-point effect) is about as strong as the influence from the EP2A itself.  

We aim to disentangle these two influences on the dynamics in the following calculation.
As the first step, we rewrite the survival amplitude in Eq. (\ref{A.surv.exp}) as
\beq
  A(t)  
	= - \lamEP^2 \frac{\partial I (\lamEP,t) }{\partial  \lamEP}  
\label{A.surv.diff}
\eeq
in which
\beq
  I (\lamEP,t)
	= \int_{C}\frac{d\lam}{2\pi i}\,\left(-\lam+\frac{1}{\lam}\right)\exp\left[i\left(\lam+\frac{1}{\lam}\right)t\right]
			\frac{1}{\lam-\lamEP}.
\label{A.surv.I}
\eeq
In this step, we have reduced the order of the pole by introducing differentiation with respect to $\lamEP$.
In the next step, we rewrite the integration into the complex energy plane through the change of integration variable $ \lambda = -E/2 + i \sqrt{1-E^2/4} $, which yields
\beq
  I (\lamEP,t) 
	= - \frac{1}{2 \pi} \int_{C_E} dE\,e^{-i E t} \frac{\sqrt{1-E^2/4}}{ E-\eEP}
	,
\label{I.int.E}
\eeq
in which $C_E$ is the counter-clockwise contour surrounding the branch cut in the complex energy plane
and $\eEP=-\lamEP-\lamEP^{-1}$.
Now, although the order of the pole has been effectively reduced, it still appears in the integration simultaneously with the branch cut $\sqrt{1-E^2/4}$.  However, following the development from App. C of Ref. \cite{GO17}, we can remove the pole entirely by rewriting it in terms of an additional integration, such that Eq. (\ref{I.int.E}) now takes the form
\beq
  I (\lamEP,t)  =-\frac{i }{\pi} \int_0^\infty d\tau\int_{-2}^2 dE\,e^{-i E t} e^{i(E-\eEP)\tau}\sqrt{1-E^2/4} 
  	.
\label{I.tau}
\eeq
The integration in $E$ can now be rewritten in terms of a Bessel function, so that Eq. (\ref{I.tau}) takes the form
\beq
 I (\lamEP,t) = e^{-i \eEP t}\left[ 1/\lamEP -i  \int_{0} ^t dt' \, e^{i \eEP t'}\frac{J_1(2t')}{t'} \right]
 	.
\label{I.J1}
\eeq
Under the integral, the information about the branch cut has now been encoded in the Bessel function $J_1 (t)$.

As detailed in Ref. \cite{GO17}, after re-inserting Eq. (\ref{I.J1}) back into Eq. (\ref{A.surv.diff}), the remaining integration can be expanded in terms of the energy gap
\beq
  \delEP
  	\equiv -2-\eEP
	.
\label{del.ep}
\eeq
between the EP2A and the band edge (lower threshold).
This expansion introduces a factor $t^{-1/2}$, which as detailed in Ref. \cite{GPSS13} (see also \cite{GNOS19}) is a direct result of the proximity of the eigenvalues to the band edge.  Under the assumption $|\delEP t|\ll 1$, or $t \ll T_{EP}$ with
\beq
  T_{EP}
  	\equiv \frac{1}{\delEP},
\label{T.EP}
\eeq
we find the approximation for the survival amplitude as
\beq \label{eq:BAsymp2}
A (t) \approx e^{-i \eEP t}\left[1 - 4 \sqrt{\frac{ i t\delEP }{\pi}} + 2 it \delEP\right] + O(t\delEP )^{3/2}
	.
\eeq
The second term here is our key result, which can be understood as arising due to the combination of the factor $t^{-1/2}$ coming from the band edge and the factor $t e^{-i \eEP t} $ coming from the double pole at the EP2A.  Out of caution, we also keep the third term to maintain consistency in keeping up with one correction term when we take the modulus squared for the survival probability.


\begin{figure}[h]
\begin{minipage}{16pc}
\includegraphics[width=16pc]{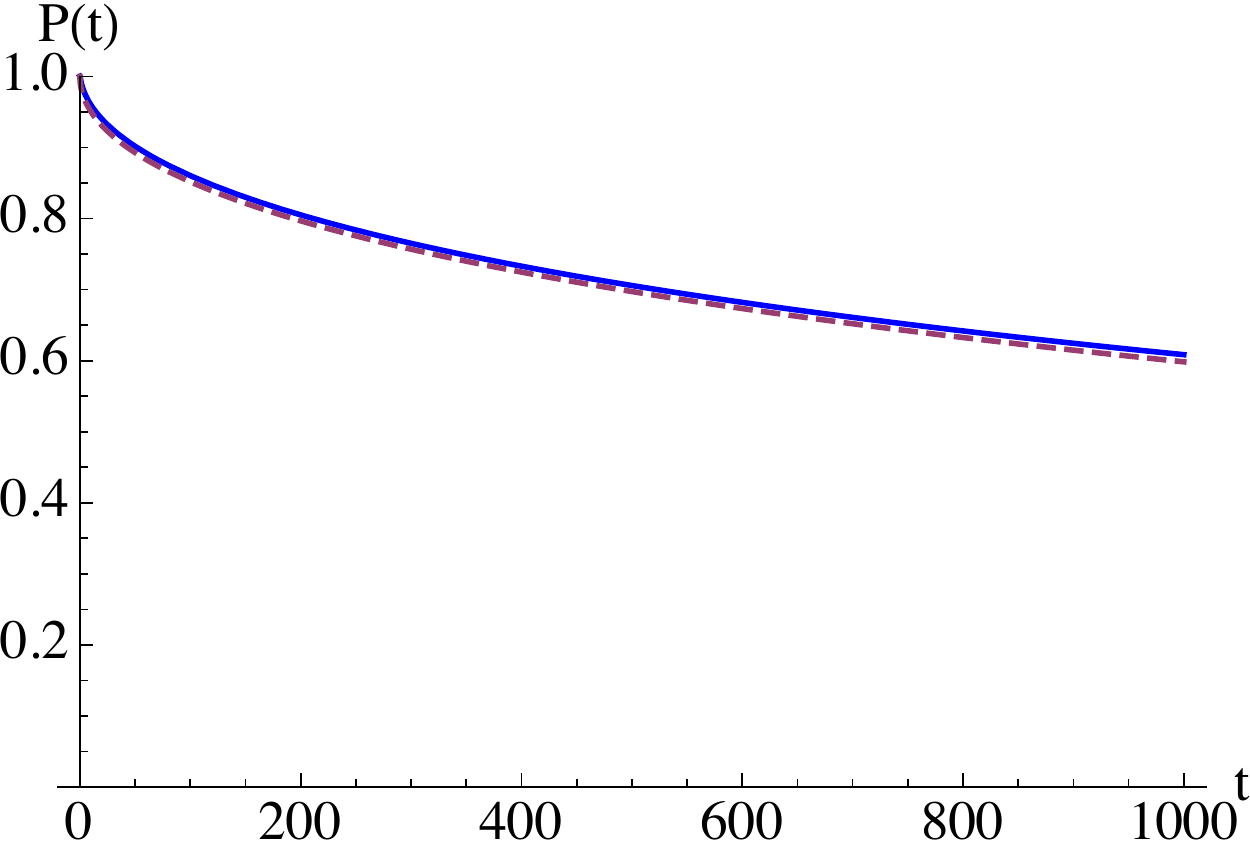}
\caption{\label{fig:EP2A.1}
Survival probability $P(t)$ near the EP2A, which in turn is near the band edge.  The solid blue curve gives the numerical integration while the purple dashed  curve represents the approximation for the intermediate timescale dynamics reported in Eq. (\ref{A.P.band.edge}).  The parameters used are $g=0.1$ and $\epsd = -1.989974$, very near the EP2A at $\epsEP \approx -1.989975$.}
\end{minipage}\hspace{2pc}%
\begin{minipage}{16pc}
\includegraphics[width=16pc]{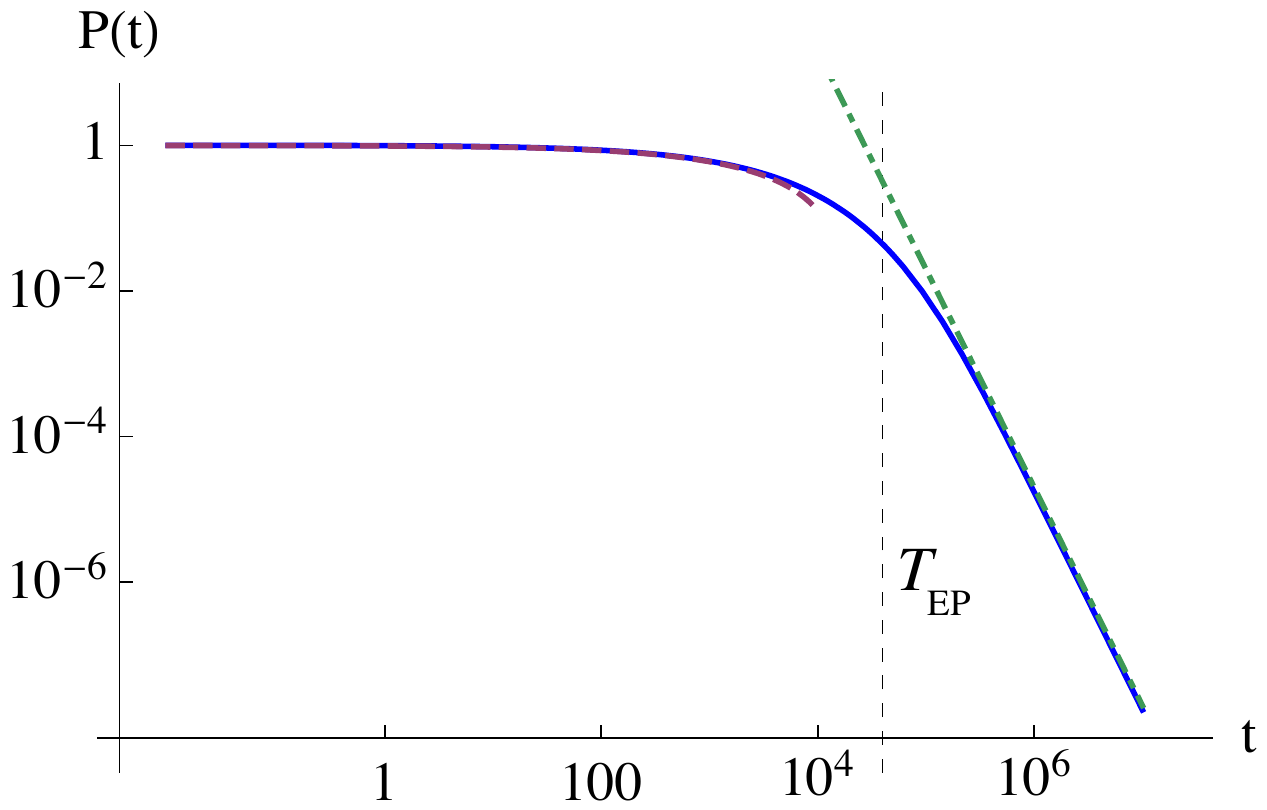}
\caption{\label{fig:EP2A.2}
Log-log plot of the survival probability near the EP2A that is in turn near the band edge.  The numerical integration (solid blue curve) is compared with the Eq. (\ref{A.P.band.edge}) approximation (purple dashed  curve) for $t \ll T_{EP}$ and with the long-time approximation Eq. (\ref{A.surv.prob.fz}) (green chained line) for $T_{EP} \ll t$.
The parameters used are the same as Fig. \ref{fig:EP2A.1}, with $T_{EP} \sim 10^5$.
}
\end{minipage} 
\end{figure}

Taking the square modulus of Eq. (\ref{eq:BAsymp2}) for the physical quantity, we
obtain the expression for the survival probability on the intermediate timescale as
\beq
  P (t)
  	\approx 1 -4 \sqrt{ \frac{2 t \delEP}{\pi}} + \frac{16 t \delEP}{\pi}
	.
\label{A.P.band.edge}
\eeq
This result is plotted (purple dashed curve) against the numerical integration (blue solid curve) in Fig. \ref{fig:EP2A.1} (linear scale) and Fig. \ref{fig:EP2A.2} (log scale) for the case $g=0.1$ and $\epsd = -1.989974$.  For this value of $g$, the EP is located $\epsEP \approx -1.989975$ and the coalesced eigenvalue is located in the energy plane at $\eEP \approx -2.0000253$, very near the band edge at $E=-2$.  Fig. \ref{fig:EP2A.2} demonstrates that Eq. (\ref{A.P.band.edge}) describes the evolution quite accurately up until about 
$t \sim T_{EP}$, where $T_{EP} = 1/\delEP  \sim 10^5$ as defined in Eq. (\ref{T.EP}).

For $t \gg T_{EP}$, the usual branch-point dynamics take over \cite{GPSS13,Muga_review,GCMM95}, yielding inverse power law decay.  For the present model, it is shown in Ref. \cite{GO17} that the long-time dynamics at the EP2A are given by
\beq
  P (t)
	\approx \frac{g^4}{4\pi \left( 1 - \sqrt{1-g^2} \right)^8 t^{3}}
	.
\label{A.surv.prob.fz}
\eeq
This result agrees well with the numerical integration for the dynamics for $t \gg T_{EP}$ as shown in Fig. \ref{fig:EP2A.2} (green chained line).

\begin{figure}[h]
\begin{minipage}{16pc}
\includegraphics[width=16pc]{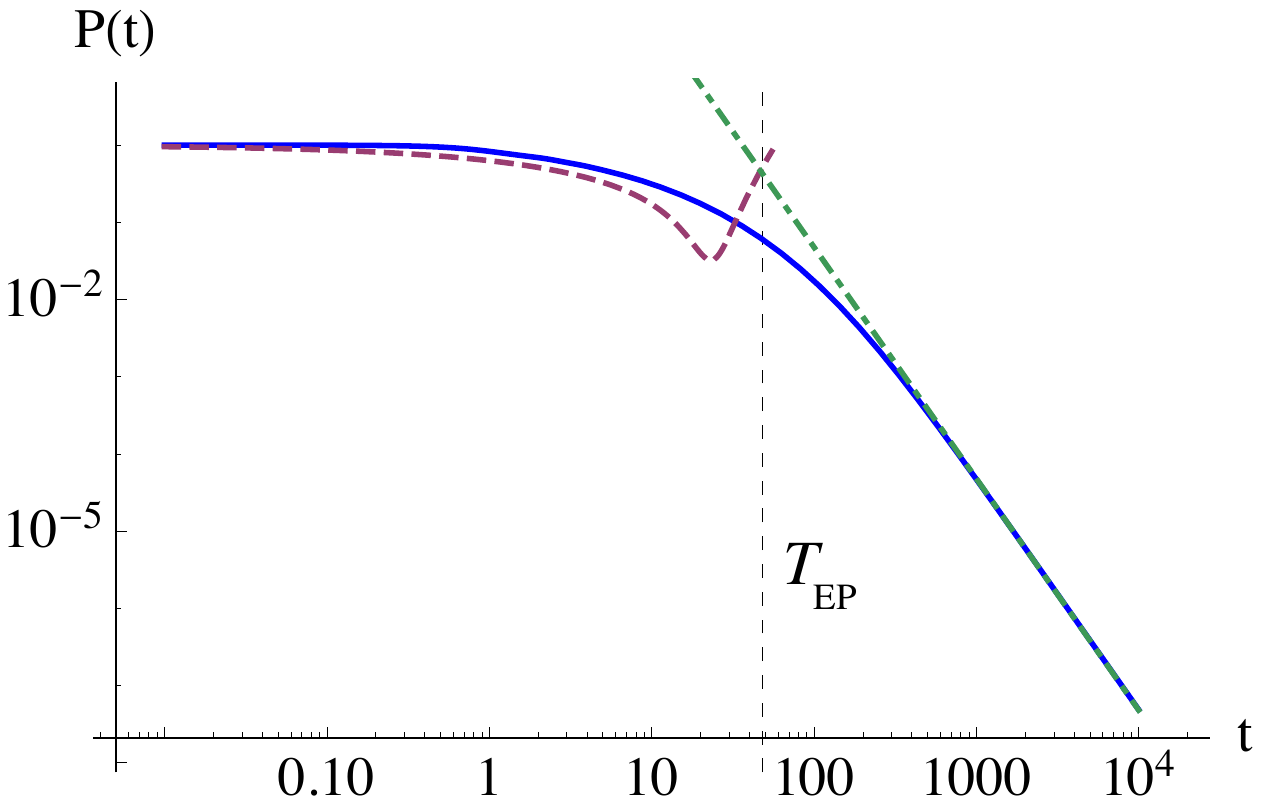}
\caption{\label{fig:EP2A.3}
Log-log plot of the survival probability $P(t)$ near the EP2A for $g=0.5$, giving $\eEP \approx -2.02073$
and $\epsEP \approx -1.73205$.  The nearby value of $\epsd$ has been chosen as $\epsd = -1.7321$.
The solid blue curve gives the numerical integration, the dashed purple curve represents the expression reported in Eq. (\ref{A.P.band.edge}) and the chained green line gives the long time evolution $\sim t^{-3}$ from  Eq. (\ref{A.surv.prob.fz}).
}
\end{minipage}\hspace{2pc}%
\begin{minipage}{16pc}
\includegraphics[width=16pc]{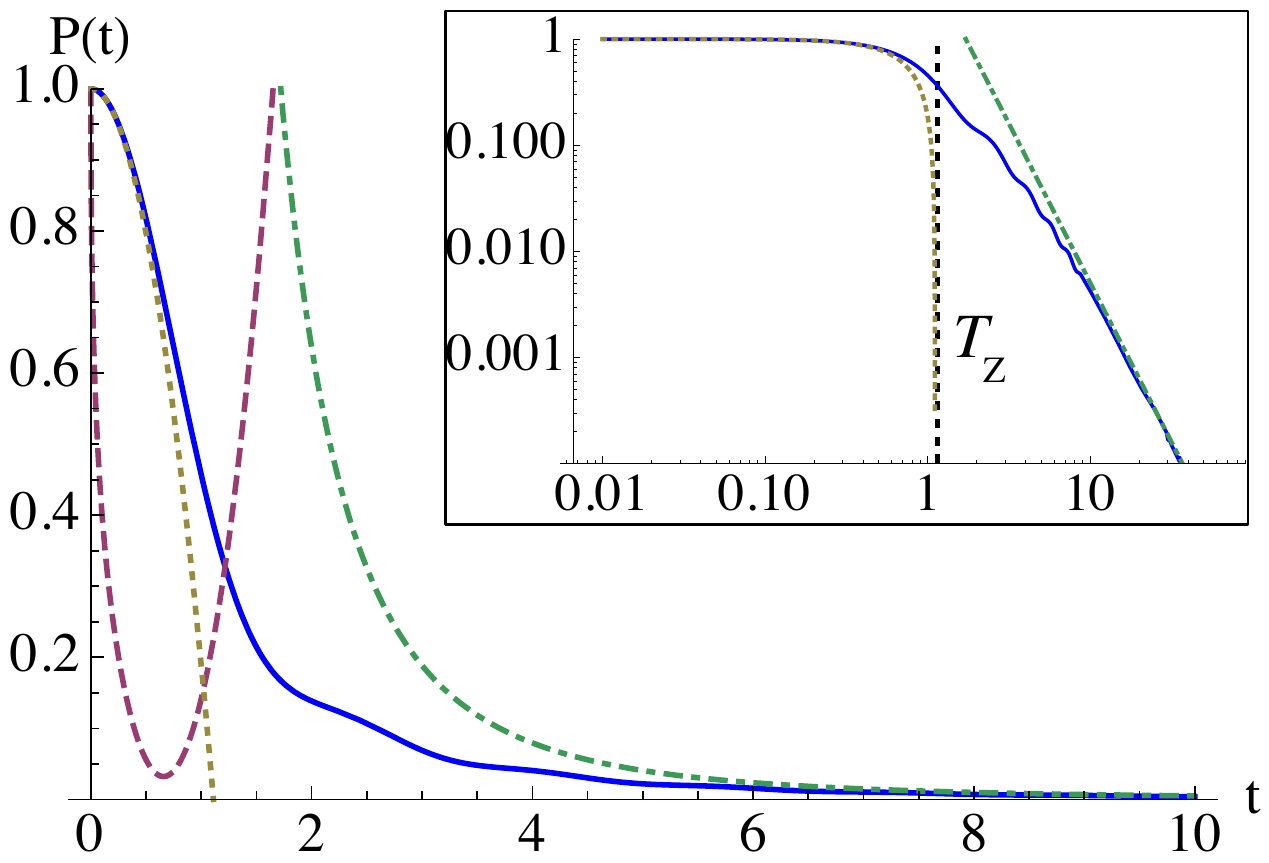}
\caption{\label{fig:EP2A.4}
Survival probability near the EP2A when it is farther from the band edge [inset: log-log plot].  Here $g=0.9$, which places the EP2A in the energy plane at $\eEP \approx -2.73005$ and in the parameter space at $\epsEP \approx -0.87178$.  We have set $\epsd = -0.87$, nearby $\epsEP$.
In this case, the short-time parabolic dynamics $P_Z (t) \approx 1 - g^2 t^2$ in Eq. (\ref{A.zeno}) (beige dotted curve) is strongly pronounced, while the prediction from  Eq. (\ref{A.P.band.edge}) [purple dashed curve] fails completely.
}
\end{minipage} 
\end{figure}

Next, we briefly comment on how the dynamics are modified as the parameters are changed so that the exceptional point is moved away from the band edge.  For the case $g=0.5$ shown in Fig. \ref{fig:EP2A.3}(a), the EP2A appears at $\eEP \approx -2.02073$, so that the transition from the $P(t) \sim 1 - C t^{1/2}$ dynamics to the long-time dynamics $P(t) \sim 1/t^3$ occurs earlier, around $T_{EP} \sim 50$.  Notice that the survival probability is also significantly less depleted when the long-time dynamics kick in, as compared to Fig. \ref{fig:EP2A.2}.

Finally, in Fig. \ref{fig:EP2A.4} we consider the case in which $g=0.9$ and $\epsd = -0.87$ so that the EP is even farther from the band edge and the timescale $T_{EP}$ now actually falls before the timescale $T_Z \approx 1.15$ for the short-time dynamics [from Eq. (\ref{T.Z})].  
Because $T_{EP} < T_Z$, the fractional power law dynamics on the intermediate time scale are now squeezed out of the picture entirely, in favor of the short-time parabolic dynamics that are greatly enhanced from the usual picture.  Further, in this interesting case, the short-time dynamics transition directly to the long-time inverse power law dynamics, which are also significantly enhanced compared to the usual picture.  This case demonstrates that the dynamics associated with an EP2A {\it far} from the band edge could offer an opportunity to study both the short-time Zeno dynamics and the long-time inverse power law dynamics that are usually quite challenging to observe in experiment.


\section{Quantum dynamics at an EP3A near the threshold}
\label{sec:EP3A}

In the previous section, we saw that the dynamics associated with an EP2A were fully non-Markovian, and that in the case that the exceptional point was located near the band edge the survival probability took the characteristic form $P(t) \sim 1 - C_1 \sqrt{t} + C_2 t$.  In this section, we present an example of an EP3A near the band edge and give a quick numerical example confirming the dynamics are again non-Markovian.  We discuss a possible fitting for our simulation as well, but working out the details of the evolution at an analytic level is left for future work.

Let us consider the Hamiltonian
\beq
  H_{n}
  	= \epsd d^\dagger d - J \sum_{j=1}^{\infty} \left( c_j^\dagger c_{j+1} + c_{j+1}^\dagger c_j \right)
		- g \left( c_n^\dagger d + d^\dagger c_n \right)
	,
\label{ham.3A}
\eeq
which consists of a discrete, semi-infinite array with a quantum emitter or dot side-coupled to the $n$th element of the array.  At least for the choices $n = 2, 4$ and $6$, the model can be shown to contain an EP3A that moves nearer to the threshold with increasing $n$ (it is likely this pattern would continue for even $n>6$).  For the choice $n=2$ the model is more analytically tractable, but the EP3A is unfortunately not so close to the band edge so that an expansion like what we employed in Sec. \ref{sec:EP2A} seems inapplicable anyways.  So instead we will mainly focus on the $n=4$ model to show the properties of the exceptional point and a numerical simulation confirming the non-Markovian dynamics.  We will then briefly discuss the $n=6$ model and give a simulation for the non-Markovian dynamics in that case as well.

The variable parameters in the model in Eq. (\ref{ham.3A}) are again the potential on the dot $\epsd$ and the dot-chain coupling $g$.  As usual, we measure the energy in units of $J=1$.  


\subsection{The $n=4$ model: spectrum and EP3A near the band edge}
\label{sec:EP3A.4}

Here we consider the $n=4$ case of the model in Eq. (\ref{ham.3A}).
Since the preceding analysis based on the effective Hamiltonian was primarily introduced for performing analytic calculations in the previous EP2 cases, we will here instead rely directly on the Green's function formalism, which is a bit quicker to dive into for present purposes.


As a first step, we can partially diagonalize $H_n$ by applying the Fourier transform on the half-chain in the form
\beq
  c_n^\dagger = \sqrt{\frac{2}{\pi}} \int_0^\pi dk \; \sin k \ \tilde{c}_k^\dagger
  	.
\label{fourier}
\eeq
This allows us to write Eq. (\ref{ham.3A}) for $n=4$ in the form
\beq
  H_{n=4}
  	= \epsd d^\dagger d + \int_0^\pi dk \ E_k \tilde{c}_k^\dagger \tilde{c}_k
		+ g \int_0^\pi V_k \left( \tilde{c}_k^\dagger d + d^\dagger \tilde{c}_k \right)
	,
\label{ham.4}
\eeq
in which $E_k = -2 \cos k$ is the usual tight-binding dispersion and $V_k = - \sqrt{2/\pi} \sin 4 k$.  
The Green's function at the side-coupled dot impurity site can be now obtained at the single-particle level, following resummation, in the form
\beq
  \bra d | \frac{1}{E - H_{n=4}} | d \ket = \frac{1}{E - \epsd - \Sigma_{n=4} (E)}
  	,
\label{3A.green}
\eeq
in which the self-energy function $\Sigma_{n=4} (E)$ is given by
\beqa
  \Sigma_{n=4}(E) \equiv g^2 \int_0^\pi \frac{| V_k |^2}{E - E_k}
   	=  \frac{g^2}{\sqrt{E^2 - 4}} \left[ 1 - \left( \frac{-E + \sqrt{E^2 - 4}}{2} \right)^8 \right]
	.
\label{3A.self}
\eeqa
The discrete eigenvalue spectrum is then determined by the poles of Eq. (\ref{3A.green}), or $E - \epsd - \Sigma_{n=4} (E) = 0$.  Applying Eq. (\ref{3A.self}) in this relation yields an octic polynomial equation $p_{n=4} (E_j) = 0$ in which
\beqa
  p_{n=4} (E) 
   &  	= & g^2 E^8 -\epsd \; g^2  E^7 -g^2 (6 + g^2) E^6 + 6 \epsd \ g^2 E^5 + 2 g^2 (5 + 2g^2) E^4
   					\nonumber	\\
   & &		- 10 \epsd \ g^2 E^3 - (1 + 2g^2)^2 E^2 + 2 \epsd (1 + 2g^2) z - \epsd^2
   	.
\eeqa

Assuming $g \ll 1$, we find that for most values of $\epsilon_d$ the spectrum consists of three resonance/antiresonance pairs and two real-valued solutions.  Approximately speaking, for values $|\epsd| \lesssim 2$ inside the band, the two real solutions are virtual states, while for $|\epsd| \gtrsim 2$ one is usually a bound state and the other is virtual.

\begin{figure}[h]
\includegraphics[width=18pc]{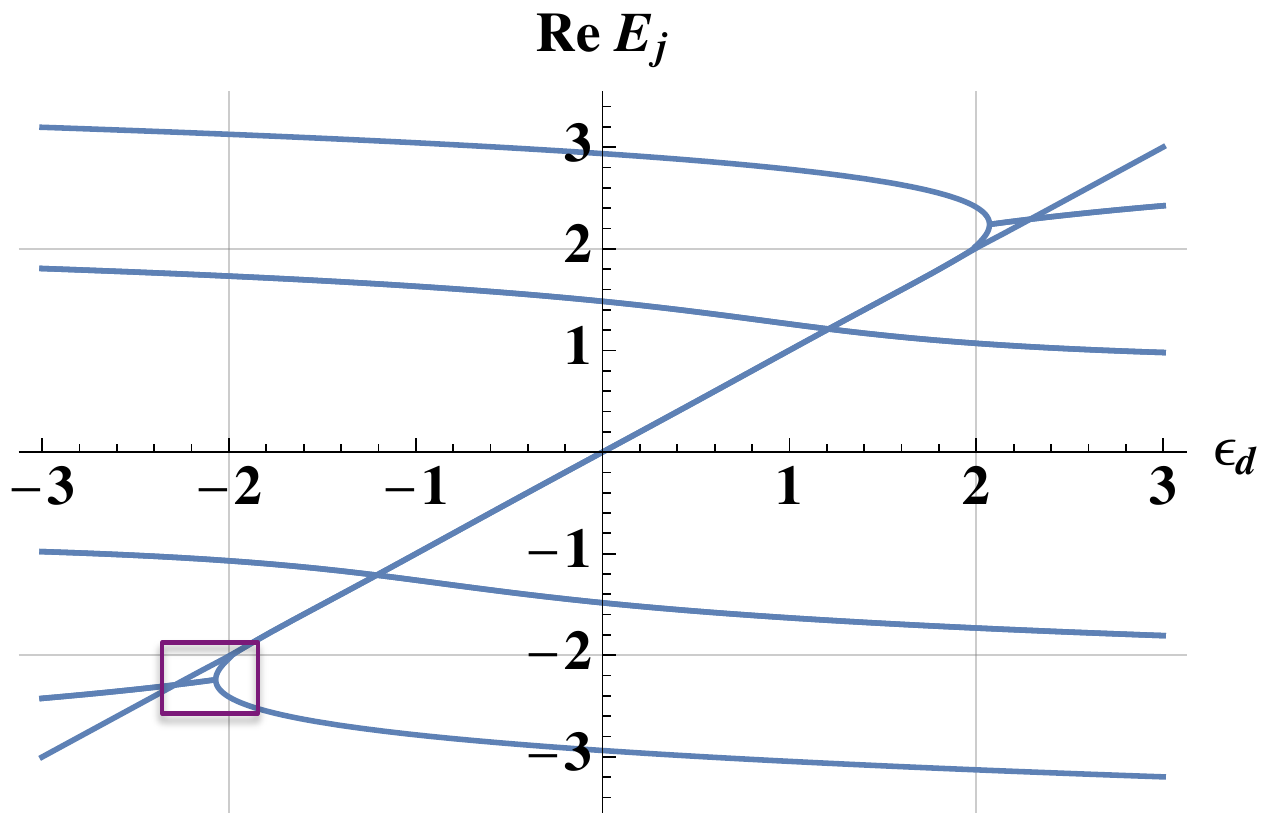}\hspace{2pc}%
\begin{minipage}[b]{14pc}
\caption{\label{fig:EP3A.1}  Real parts of the discrete eigenvalues for model $H_{n=4}$ as a function of $\epsd$ in the case $g=0.06$.  The upper and lower band edges $\pm 2$ are indicated by grey lines on both axes.  The area highlighted by a purple box around the lower band edge is given a closer view in Fig \ref{fig:EP3A.2}.
}
\end{minipage}
\end{figure}

\begin{figure}[h]
\begin{minipage}{16pc}
\includegraphics[width=16pc]{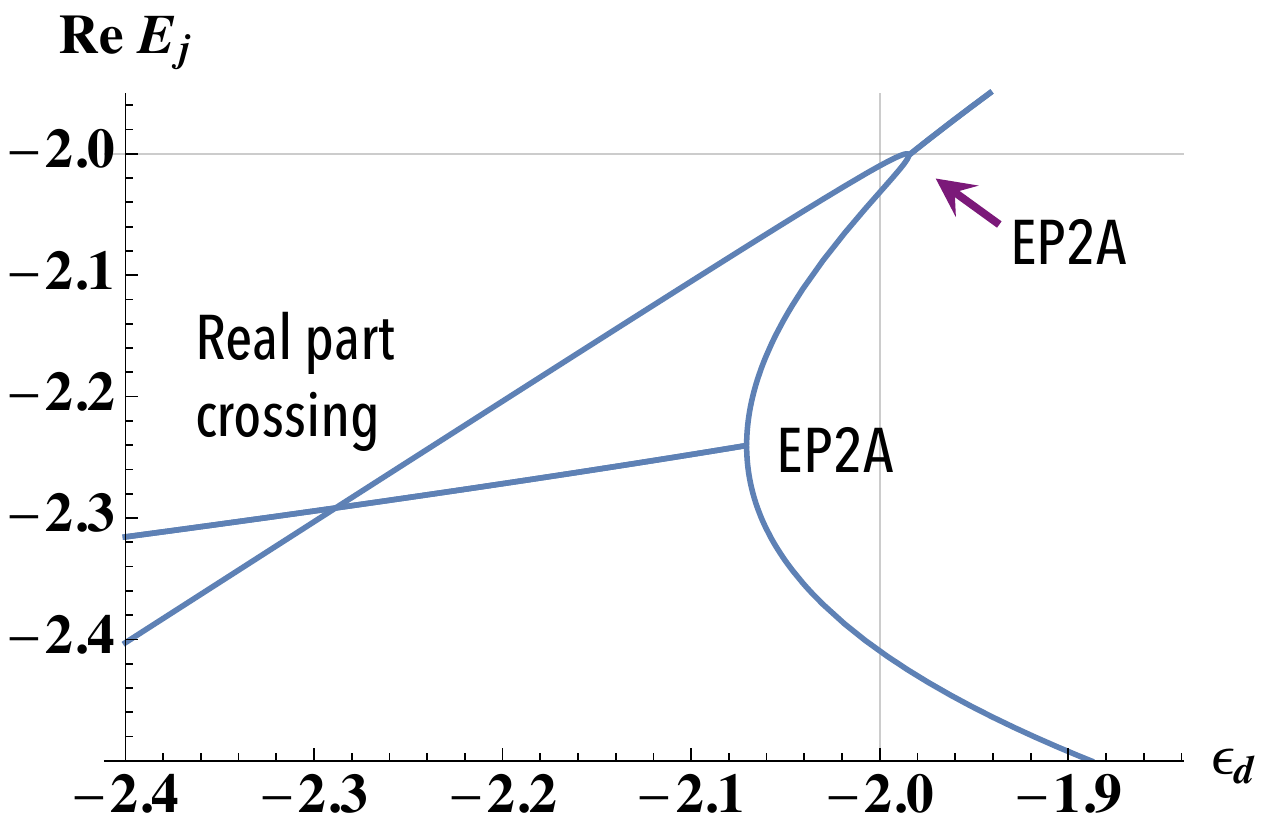}
\caption{\label{fig:EP3A.2}
Zoomed-in view of the real part of the spectrum for $H_{n=4}$ corresponding to the purple box from Fig. \ref{fig:EP3A.1} (for $g=0.06$).  The two EP2As near the lower band edge are indicated, as well as a crossing for the real parts of the eigenvalue.
}
\end{minipage}\hspace{2pc}%
\begin{minipage}{16pc}
\includegraphics[width=16pc]{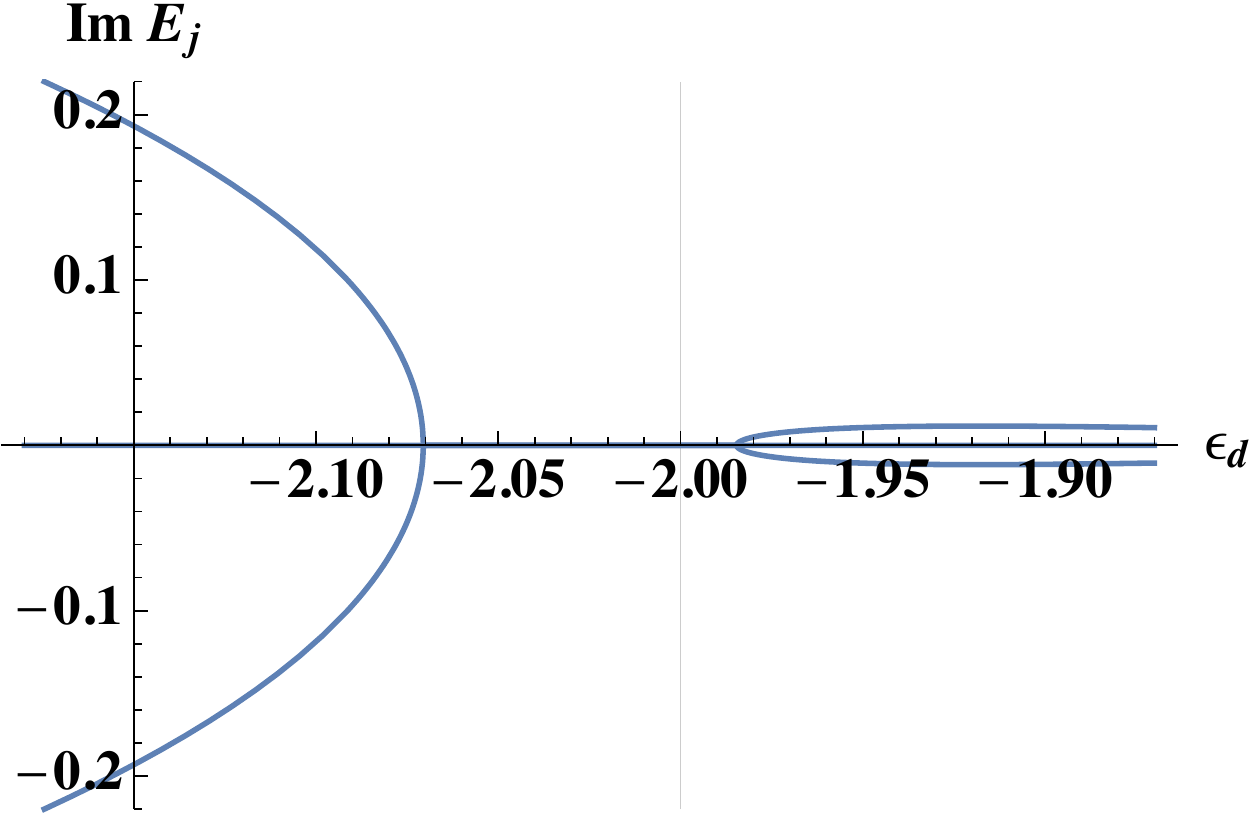}
\caption{\label{fig:EP3A.3}
Imaginary part of the spectrum for $H_{n=4}$ showing the meeting points for two resonance/anti-resonance pairs, corresponding to the two EP2As from Fig. \ref{fig:EP3A.2} (for $g=0.06$).  Note that the purely real eigenvalues also appear along the horizontal axis.
}
\end{minipage} 
\end{figure}

In Fig. \ref{fig:EP3A.1}, we plot the real part of the eigenvalues for the representative case $g=0.06$.  Notice that near the upper and lower band edges, a region occurs in which several eigenvalue crossings appear.  The region near the lower band edge, which is highlighted with a purple box in Fig. \ref{fig:EP3A.1}, can be seen in a zoomed-in view in Fig. \ref{fig:EP3A.2}.  In this region, there are two EP2As and the spectral properties depend sensitively on the precise value of $\epsd$.  Around $\epsd \approx -2.3$ there occurs a crossing of the real part of the eigenvalues between a bound state and a resonance/anti-resonance pair.  We emphasize this point is not an exceptional point.  However, as we sweep the value of $\epsd$ leftward in the figure, we encounter an EP2A around $\epsd \approx -2.07$ at which the previously-mentioned resonance/anti-resonance pair coalesces before forming two virtual states (or two anti-bound states \cite{HO14}).  The virtual state with the more negative eigenvalue very slowly moves off to negative infinity for increasing $\epsd$, while the other instead approaches a third virtual state very near the lower band edge $-2$.  These latter two virtual states coalesce to form a new resonance/anti-resonance pair around $\epsd \approx -1.985$.

Next, in Fig. \ref{fig:EP3A.3}, we show the imaginary part of the complex eigenvalues, zooming-in a bit more tightly on the region in which the two EP2As occur.  The location of the EP2As at which the negative and positive parts of the imaginary eigenvalues for the resonance and anti-resonance collapse is clearly visible.   We emphasize there are three virtual states (all three with real eigenvalue) in the region between the EP2As, which appear along the horizontal axis in Fig. \ref{fig:EP3A.3}. 

\begin{figure}[h]
\begin{minipage}{16pc}
\includegraphics[width=16pc]{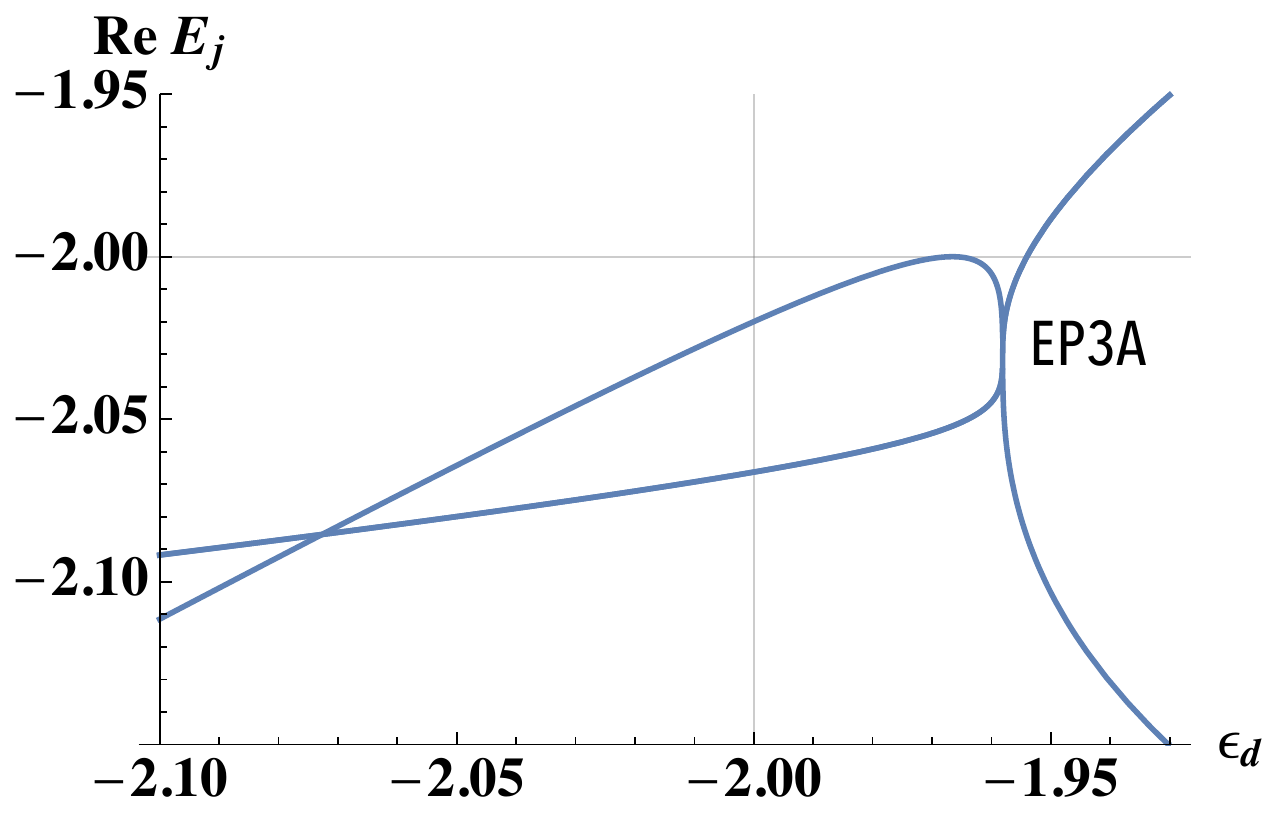}
\caption{\label{fig:EP3A.4}
Real part of the eigenvalues for model $H_{n=4}$ near the lower band edge for $g = \bar{g}_{EP3} \approx 0.0914264$.  The two EP2As from Fig. \ref{fig:EP3A.2} and Fig. \ref{fig:EP3A.3} have combined to form an EP3A at $\epsd = \bar{\epsilon}_{EP3} \approx -1.958109$.
}
\end{minipage}\hspace{2pc}%
\begin{minipage}{16pc}
\includegraphics[width=16pc]{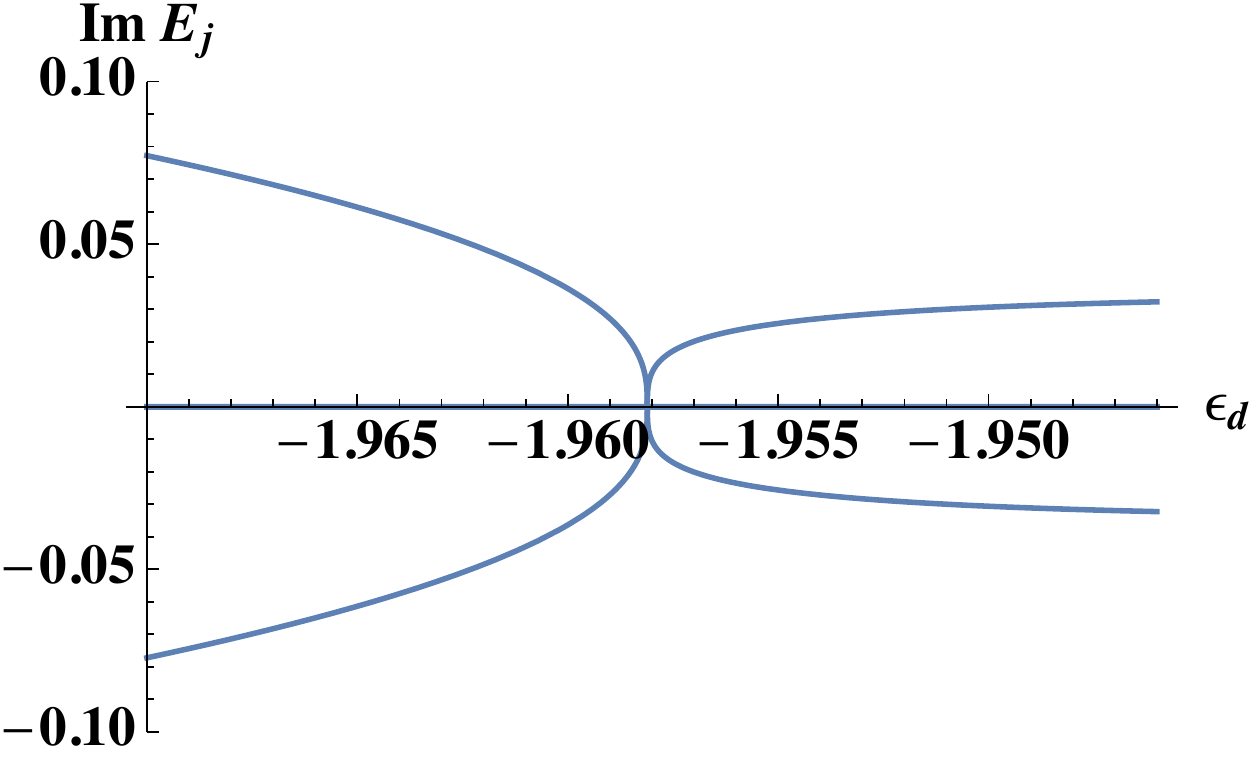}
\caption{\label{fig:EP3A.5}
Imaginary part of the eigenvalues for model $H_{n=4}$ near the lower band edge for $g = \bar{g}_{EP3} \approx 0.0914264$.  The two EP2As from Fig. \ref{fig:EP3A.2} and Fig. \ref{fig:EP3A.3} have combined to form an EP3A at $\epsd = \bar{\epsilon}_{EP3} \approx -1.958109$.
}
\end{minipage} 
\end{figure}

From this point, if we slowly increase the value of $g$ we find that the two EP2As from Figs. \ref{fig:EP3A.2} and \ref{fig:EP3A.3} gradually approach one another along the $\epsd$ axis.  When $g$ reaches the value $g = \bar{g}_{EP3} \approx 0.0914264$ the two EP2As meet at the $\epsd$ value $\epsd = \bar{\epsilon}_{EP3} \approx -1.958109$ and form an EP3.  If we fix $g$ at exactly $g = \bar{g}_{EP3}$ and vary $\epsd$ in the vicinity of the EP3, we find that a virtual state, a resonance, and an anti-resonance all coalesce at $\epsd = \bar{\epsilon}_{EP3}$, before forming another virtual state, resonance, anti-resonance trio on the other side of the EP3, as shown in Figs. \ref{fig:EP3A.4} and \ref{fig:EP3A.5}.   We refer to this type of exceptional point as an EP3A.  The coalesced eigenvalue at the EP3A is given by  $\bar{E}_{EP3} \approx -2.030646$.

We note that the above picture in which the two EP2As form an arc in the $\epsd, g$ parameter space of $H_{n=4}$ before colliding to form a higher-order exceptional point can be viewed as an {\it exceptional nexus} in the language of Ref. \cite{ENexus}.


\subsection{The $n=4$ model: EP3A dynamics}
\label{sec:EP3A.dynamics}

We note that the exceptional point eigenvalue $\bar{E}_{EP3} \approx -2.030646$ appears somewhat close to the lower band edge (threshold) at $E = -2$, so that the characteristic gap $\Delta_{EP3}$ takes the value $\Delta_{EP3} = -2 - \bar{E}_{EP3} \approx 0.030646$.   Hence we expect that the decay will be fully non-Markovian and further that the threshold should have some significant influence on the dynamics in combination with the exceptional point itself.

\begin{figure}[h]
\begin{minipage}{16pc}
\includegraphics[width=16pc]{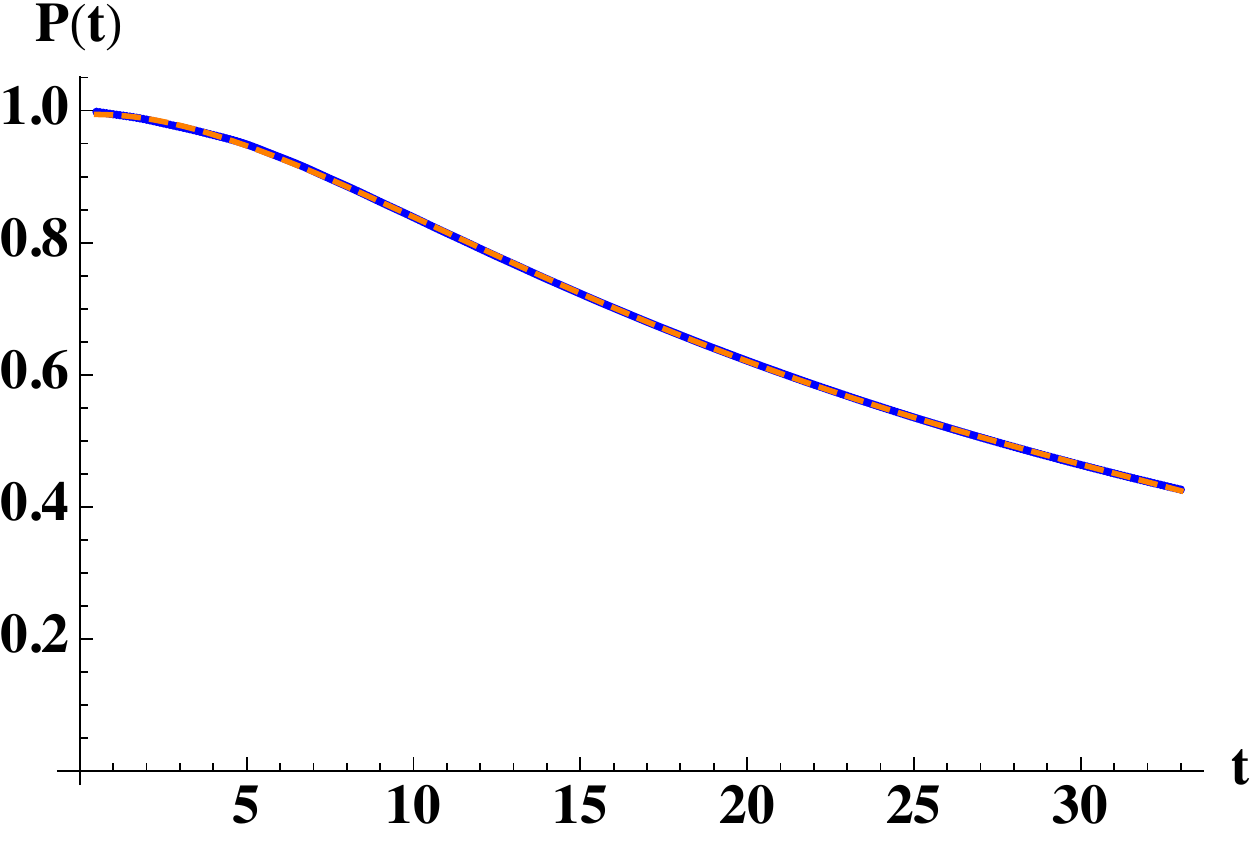}
\caption{\label{fig:n4.lin}
Numerical simulation (blue data points) for the survival probability $P(t)$ near the EP3A in model $H_{n=4}$.  The parameters used are $\epsd = -2.030647$ and $g=0.0914264$.
The data fit for Eq. (\ref{EP3A.P}) is shown as the orange dashed line with the parameters reported in Table \ref{coeffs} for $n=4$.
}
\end{minipage}\hspace{2pc}%
\begin{minipage}{16pc}
\includegraphics[width=16pc]{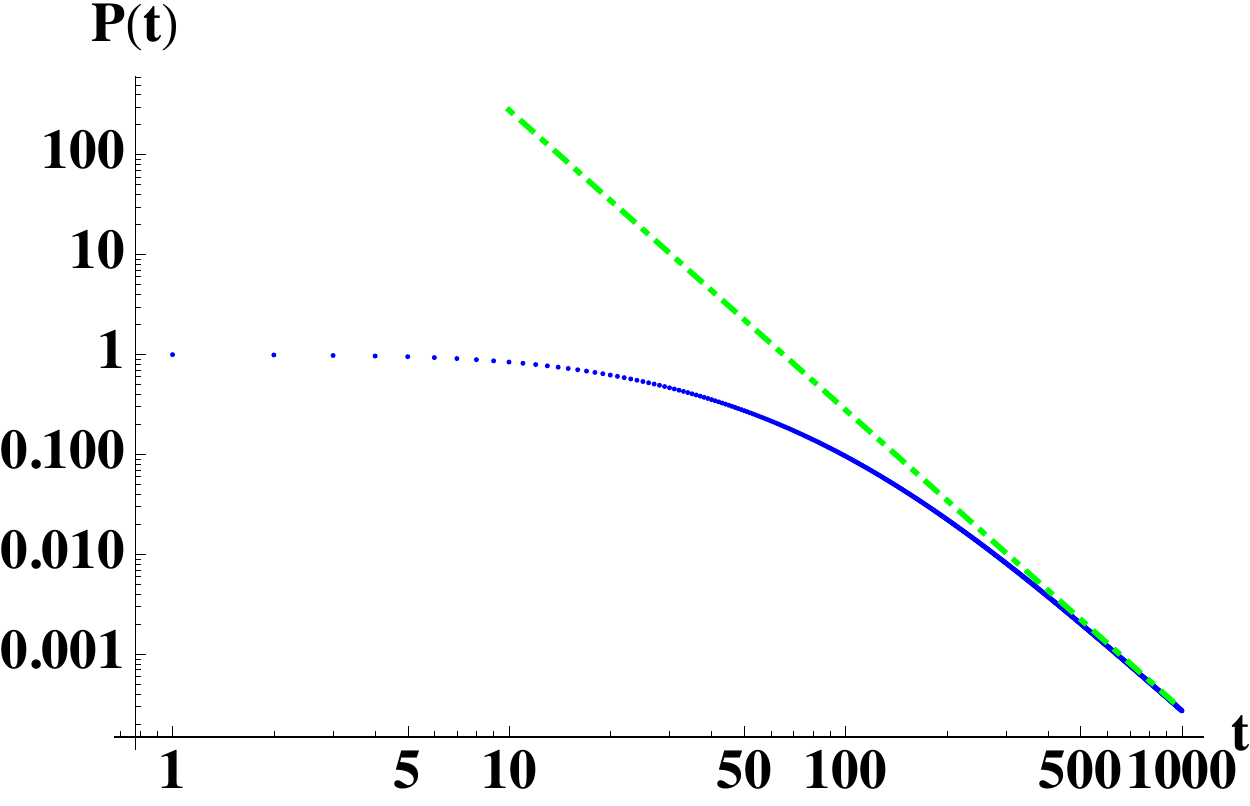}
\caption{\label{fig:n4.log}
Log-log plot of the numerical simulation (blue data points) for the survival probability $P(t)$ near the EP3A in model $H_{n=4}$.  The parameters used are $\epsd = -2.030647$ and $g=0.0914264$.
The analytic expression for the long-time dynamics from Eq. (\ref{EP3A.lt}) is shown with the green dot-dashed line.
}
\end{minipage} 
\end{figure}

We plot a numerical simulation for the dynamics in the case $\epsd = -2.030647$ and $g=0.0914264$ (very close to the EP3A) as the blue data points in Fig. \ref{fig:n4.lin}.  
This numerical result was obtained by direct solution of Schr\"odinger's equation for a model with 1050 lattice sites beyond the $n$th site (with $n=4$ in the present case).  Visually, it is easy to ascertain that the decay is non-Markovian in this figure.  To be certain, we applied an exponential data fitting (and a few variations on exponential decay), which fit the data poorly.  

We attempt to better understand the non-Markovian dynamics as follows.
Working by analogy with the results for the EP2A near the band edge in Sec. \ref{sec:EP2A}, one might expect the survival amplitude in the present case to take the form
\beq
  A(t) \sim \left( 1 + B_1 t^{1/2} + B_2 t  + B_3 t^{3/2} 
			+ B_4 t^2 + B_5 t^{5/2} + B_6 t^3 \right) e^{- i \bar{E}_{EP3} t}
	,
\label{EP3A.A}
\eeq
in which the $t^{3/2}$-dependent term is expected to appear due to the multiplication of a factor $t^{2} e^{-i \bar{E}_{EP3} t}$ from the triple-pole at the EP3A and a factor $1/\sqrt{t}$ coming from the nearby band edge.
This would correspond to Eq. (\ref{eq:BAsymp2}) for the EP2A case in Sec. \ref{sec:EP2A.dynamics}, and just as in that case, we would expect it might be valid on the scale $t \lesssim T_{EP3}$, in which the timescale $T_{EP3}$ is given by
\beq
  T_{EP3} = \frac{1}{\left| -2 - \bar{E}_{EP3} \right|} \approx 32.6306
  	.
\label{T.EP3}
\eeq
Also notice that just as in the EP2A case, we have kept the higher-order $t^{m}$ terms with $m > 3/2$ in Eq. (\ref{EP3A.A}) to ensure consistency when we take the square modulus to obtain the probability as
\beq
  P(t) \sim 1 + C_1 t^{1/2} + C_2 t  + C_3 t^{3/2} 
			+ C_4 t^2 + C_5 t^{5/2} + C_6 t^3
	.
\label{EP3A.P}
\eeq
Let us emphasize that we would expect the higher-order terms to be correction terms with the $C_4$, $C_5$ and $C_6$ coefficients all being relatively small.

Assuming this picture is correct, a question immediately arises: would we expect the first ($t^{1/2}$) or second  ($t^{3/2}$) fractional power term to be the dominant term in the evolution?  If the coefficients took the comparable form as Eq. (\ref{A.P.band.edge}) for the EP2A case, then we would have $C_n t^{n/2}= A_n (\Delta_{EP3} t)^{n/2}$ appearing as powers of the small quantity $\sqrt{\Delta_{EP3} t}$.  Then $C_3 t^{3/2}$ could only be expected to be the dominant term with a much larger coefficient $A_3$ than $A_1$.

\begin{table}[h]
\caption{\label{coeffs}Fitting parameters for the hypothesized power-law decay near the EP3A [Eq. (\ref{EP3A.P})], as determined in Mathematica} 

\begin{center}
\lineup
\begin{tabular}{*{7}{l}}
\br                              
Model&$\m\m C_1$  &\m\ $C_2$&\m\ $C_3$&\m\m $C_4$  & \m\m $C_5$&$\m\m C_6$\cr 
\mr
$n=4$ &\-0.0217969\0  &0.0286801  &\-0.0122409  &\-0.00225201  
									&0.000980366 &\-0.000076716\cr
$n=6$  &\-0.011098\0&0.0101236\0  &\-0.0029150 \0  &\-0.00050665 \0 &0.000128822 & \-0.000006557\cr 
\br
\end{tabular}
\end{center}
\end{table}

We performed several data fittings based on Eq. (\ref{EP3A.P}) in Mathematica, one of which is shown by the orange dashed curve in Fig. \ref{fig:n4.lin}.  The coefficients obtained for this data fitting are shown in Table \ref{coeffs}.  The fitting agrees with the numerical simulation well and the numbers reported for the coefficients seem intuitively reasonable, with the first three coefficients $C_{1,2,3}$ all being about the order of the gap 
$\Delta_{EP3}  \approx 0.030646$, and the coefficients for the three higher-order terms decreasing in magnitude from there.  The resulting RMS value for this fitting was also reasonable with RMS $\approx 0.00105$.

However, the fitting coefficients Mathematica chose were somewhat sensitive to the time length of the simulation.  Reducing the simulation time length eventually results in the sign designations of the $C_{1,2,3}$ coefficients being flipped, which to us seemed qualitatively a little different than Eq. (\ref{A.P.band.edge}). And in any case, we found that using a polynomial fit with only integer powers of $t$
with the same number of fitting parameters as Eq. (\ref{EP3A.P}) yielded a comparable RMS value.  Hence, we only claim that the hypothesis in Eq. (\ref{EP3A.P}) {\it might} be correct and should be investigated further.

Regardless of the applicability of Eq. (\ref{EP3A.P}), the primary point is that the dynamics are fully non-Markovian.  And for the longest timescale in the system the dynamics near the EP3A can be shown to demonstrate the usual inverse power law decay.  In particular, we derived the approximation for the evolution during this time domain as
\beq
  P(t) \approx  \frac{64 \bar{g}_{EP3}^4 }
  		{\pi (2 + \bar{\epsilon}_{EP3} -4\bar{g}_{EP3}^2)^4 t^{3} }
	,
\label{EP3A.lt}
\eeq
which is shown as the green dot-dashed line in the log-log plot in Fig. \ref{fig:n4.log}.


\subsection{The $n=6$ model: EP3A dynamics closer to the threshold}
\label{sec:EP3A.dynamics.n6}

Finally, we briefly consider the EP3A in the $n=6$ model from Eq. (\ref{ham.3A}).  
Following analysis similar to that which was applied for the $n=4$ model, we can show that the $n=6$ case gives rise to a dispersion polynomial $p_{n=6} (E_j) = 0$ in which  $p_{n=6} (E)$ is 12th-order in $E$.  The spectrum for $g \ll 1$ is qualitatively similar to that for $n=4$, except that here it usually consists of five resonance/anti-resonance pairs and two real eigenvalues. Again, the exception to this statement is near the two band edges where a picture very similar to that shown in Figs. \ref{fig:EP3A.2} and  \ref{fig:EP3A.3} occurs.  And similar to that case, we find that two EP2As meet to form an EP3A, in this case occurring at   
$\bar{g}_{EP3} \approx 0.04946448$ and $\epsd = \bar{\epsilon}_{EP3} \approx -1.9816623$.  However, the coalesced eigenvalue in this case occurs a bit closer to the threshold, at $\bar{E}_{EP3} \approx -2.0131867$.  
With this value, we would expect the timescale for the validity of the proposed approximation in Eq. (\ref{EP3A.P}) to be about
\beq
  T_{EP3} = \frac{1}{\left| -2 - \bar{E}_{EP3} \right|} \approx 75.8
  	.
\label{T.EP3.n6}
\eeq

\begin{figure}[h]
\includegraphics[width=16pc]{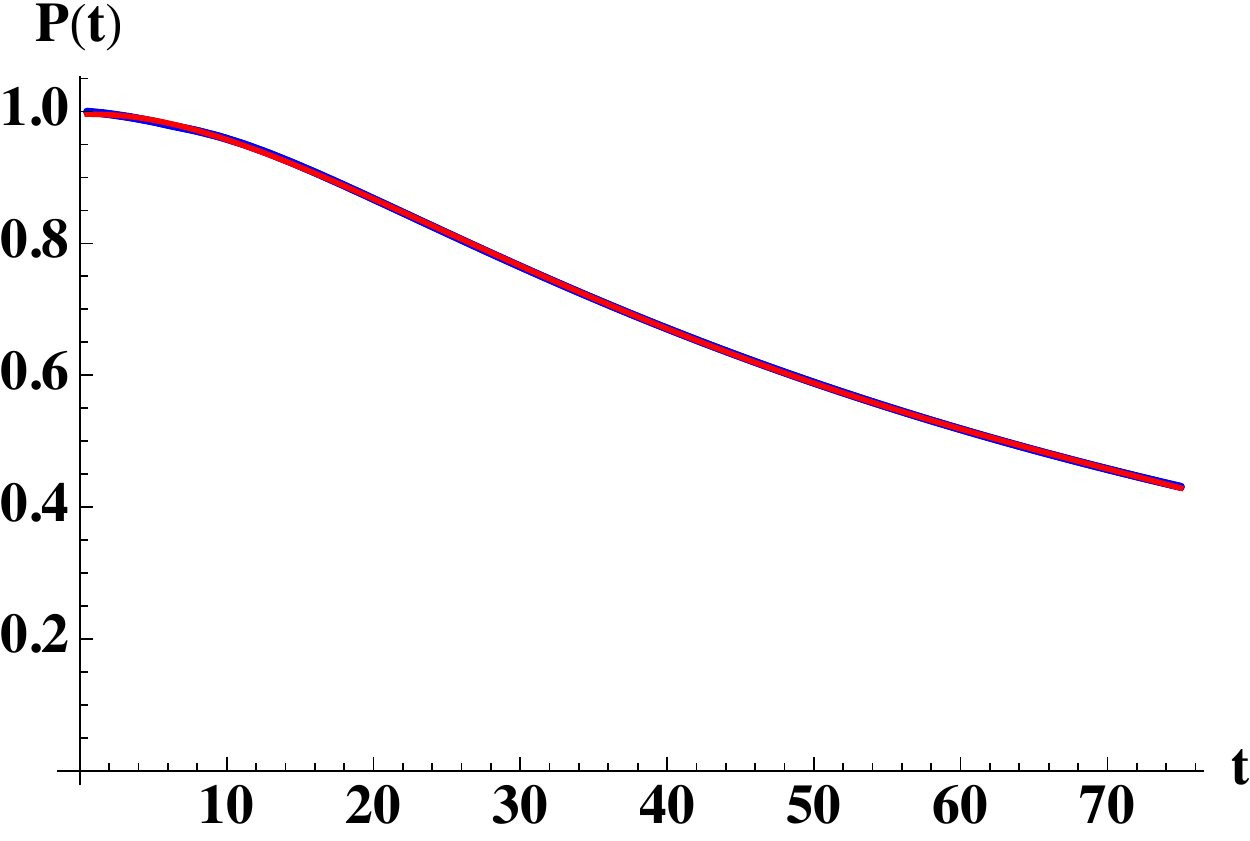}\hspace{2pc}%
\begin{minipage}[b]{16pc}
\caption{\label{fig:n6.lin}  
Numerical simulation (blue data points) for the survival probability $P(t)$ near the EP3A in model $H_{n=6}$.  The parameters used are $\epsd = -1.981662$ and $g=0.0494644$.
The data fit for Eq. (\ref{EP3A.P}) is shown as the orange line with the parameters reported in Table \ref{coeffs} for $n=6$.  The RMS value for the fitting was $0.000996$.
}
\end{minipage}
\end{figure}

We show a numerical simulation (blue data points) in Fig. \ref{fig:n6.lin} for the case $\epsd = -1.981662$ and $g=0.0494644$, very close to the EP3A.  The fitted curve corresponding to Eq. (\ref{EP3A.P}) is shown in orange, with the fitting parameters reported again in the $n=6$ line of Table \ref{coeffs}.  The estimated coefficients $C_i$ again are decreasing for the higher-order terms starting from $i=4$, although here, closer to the threshold, they are notably a bit smaller than the $n=4$ case.  It is notable that in both the $n=4$ and $n=6$ cases, the $C_1$ and $C_2$ terms seem to be a bit more dominant than the $C_3$ term.


\section{Conclusion}
\label{sec:conc}

In this work, we have reviewed coalescing eigenstates and their influence on certain physical quantities in the near-vicinity of the exceptional point at which the coalescence occurs.  In particular, we have focused on their signature influence on dynamical processes in quantum systems, taking into account that the presence of the continuum threshold can equally play a role in shaping these effects in some circumstances.

Near the B-type exceptional points, at which two or more resonances coalesce, we saw that the usual exponential decay was modified with power law factors determined by the number of coalescing resonances.  
Due to the influence of the threshold, the power law-exponential decay can be expected to be replaced by some form of non-exponential decay on very long timescales; for the EP2B considered in Sec. \ref{sec:EP2B.long}, we showed this took the form of inverse power-law decay, as plotted in Fig. \ref{fig:EP2B.2}.
We emphasize that in most typical circumstances this long-time effect would be very difficult to detect, just as in the ordinary case involving a lone resonance.
However, just as the non-exponential dynamics can be enhanced when a lone resonance approaches the threshold \cite{Jittoh05,GCV06}, it seems likely that a B-type exceptional point near the threshold would also lead to some kind of enhanced non-exponential evolution, which could prove interesting.
However, we don't know of any specific physical model in which this scenario actually occurs.

Since a coalescing resonance/anti-resonance pair at an EP2A naturally occurs in the vicinity of the continuum threshold, we found that the threshold has a much stronger influence on the dynamics near the exceptional point in this case.
As originally shown in Ref. \cite{GO17}, this results in a significantly more complex picture for the quantum dynamics.   In the case that the EP2A occurs close to the threshold, we found that the survival probability exhibited a decay of the form $P(t) \sim 1 - C_1 \sqrt{t} + C_2 t$, in which the $\sim \sqrt{t}$ term results from one factor of $t$ coming from the pole at the exceptional point and a factor $1/\sqrt{t}$ coming from the threshold.  Further, the decay was fully non-Markovian on all timescales.  In the case that the EP2A is located further away from the threshold the characteristic $1 - C \sqrt{t}$ effect becomes less pronounced, but the decay is still fully non-Markovian as the short time parabolic decay and the long-time inverse power law decay become more strongly enhanced.  This indicates that exceptional points could play a useful role in non-Markovian dynamical engineering.

We then gave an example of a model with an EP3A at which a resonance, anti-resonance and virtual state all coalesce somewhat close to the threshold.  A quick numerical simulation revealed that the dynamics were non-Markovian in this case as well.

Generalizing from these observations, another natural question to consider would be the possibility of the A-type EP appearing directly at the threshold.  Indeed, this scenario is possible in open quantum systems, and some examples have appeared in the literature \cite{TGKP16,GOH21,Heiss_zero}.  However, the detailed analysis of the time evolution in these scenarios seems relatively unexplored, with the exception of a work-in-progress by two of the present authors.  

In Ref. \cite{GOH21}, which was presented in the present virtual conference series on Pseudo-Hermitian Hamiltonians in Quantum Physics, we have demonstrated the occurrence of an exceptional point directly at the threshold in generic 1-D continuum systems that is anomalous in the sense that the branch cut appearing in the Puiseux expansion is higher-order than the number of levels that actually coalesce at the EP.  In particular, for the model studied in Ref. \cite{GOH21}, there are three levels with energy eigenvalues that converge on the continuum threshold as the system coupling $g$ is shut off: a resonance, an anti-resonance and a bound state.  This results in a Puiseux expansion for the energy eigenvalues of the form $E_j \approx E_{th} - C_j g^{4/3}$ in which $E_{th}$ is the energy corresponding to the threshold and $C_j$ is a constant, as usual.
From our previous experience, this seems to strongly suggest the presence of a third-order exceptional point at the threshold.  However, a careful analysis reveals the order of the EP is actually only second-order, as only two linear combinations of the three discrete states can be shown to coalesce, while a third linear combination instead fuses to the continuum threshold itself.  Meanwhile, the dynamics associated with this anomalous-order exceptional point takes the form $1 - C t^{3/2}$, which seems more similar to what one might expect for a third-order EP rather than a second-order one.  This result suggests that at least for the EPs appearing directly at the threshold, the time evolution probably has to be worked out on a case-by-case basis.  We also note the dynamics in this case are related to a well-known problem in spontaneous emission near a photonic band gap  \cite{LNNB00,KKS94,John94}.


We conclude with a final comment noting that in this work we have confined our analysis of the role of exceptional points in open quantum systems to the level of the Hamiltonian.  However, it has been emphasized in a number of recent works that a deeper understanding of the role of exceptional points in quantum systems can be obtained by evaluating the problem at the level of the density matrix under the Liouvillian master equation formalism, including Lindblad terms that can account for decoherence and quantum jump processes \cite{BreuPet,HatanoLindblad,LEP,Haack}.  In Ref. \cite{LEP} it is has been shown by F. Minganti, et al that exceptional points appearing at the level of the Hamiltonian can have their locations shifted under the Liouvillian formalism, and further that new exceptional points may appear that cannot be seen in the Hamiltonian picture at all.

Given the possibility for parameter shifting of EPs under the Liouvillian framework, several interesting questions immediately arise: would the aforementioned exceptional points occurring directly at the threshold also appear shifted?  And further, for exceptional points either at or near the threshold, how would the dynamics at the Hamiltonian level be modified under the influence of quantum jumps?

However, this does not necessarily imply that it would be impossible to observe the dynamics associated with the bare Hamiltonian itself.  We note two recent experiments \cite{Murch19,Murch21} conducted at Washington University by the group of K. Murch, in which data post-selection is used to eliminate experimental trials in which quantum jumps occur.  This approach has enabled studies of quantum state tomography and dynamics near an exceptional point in a superconducting qubit embedded in a cavity \cite{Murch19}.

\ack
We thank Kater Murch for helpful discussions related to this work.
S. G. acknowledges support from the Japan Society for the Promotion of Science through KAKENHI Grant No. JP18K03466. 


\end{document}